\documentclass[prx,twocolumn,secnumarabic,balancelastpage,amsmath,amssymb,noeprint,superscriptaddress]{revtex4-2}

\usepackage{graphicx}
\usepackage{natbib}
\usepackage{subcaption}
\usepackage{xcolor}
\usepackage{bm}
\usepackage{bbm}
\usepackage{physics}
\usepackage{amsmath}
\usepackage{amsthm}
\usepackage{dsfont}
\usepackage{makecell}
\usepackage{dcolumn}
\usepackage[colorlinks,citecolor=blue,linkcolor=blue,urlcolor=blue,bookmarks=false,hypertexnames=true]{hyperref}
\usepackage[mathscr]{eucal}
\usepackage{wrapfig}
\usepackage{tikz}
\usepackage{quantikz}
\usepackage{ragged2e}

\newcommand{\jb}[1]{#1}

\begin{document}
\title{
Enhancing Optical Imaging via Quantum Computation}
\author{Aleksandr Mokeev}
\affiliation{Department of Electrical Engineering, Eindhoven University of Technology, Eindhoven, 5600 MB, The Netherlands}
\author{Babak Saif}
\affiliation{NASA Goddard Space Flight Center, 8800 Greenbelt Rd, Greenbelt, MD 20771, USA}
\author{Mikhail D. Lukin}
\affiliation{Department of Physics, Harvard University, Cambridge, MA 02138, USA}
\author{Johannes Borregaard}
\email{borregaard@fas.harvard.edu}
\affiliation{Department of Physics, Harvard University, Cambridge, MA 02138, USA}

\date{\today}

\begin{abstract}
Extracting information from weak optical signals is a critical challenge across a broad range of technologies. Conventional imaging techniques, constrained to integrating over detected signals and classical post-processing, are   limited in signal-to-noise ratio (SNR) from 
shot noise accumulation in the post-processing algorithms. We show that these limitations can be circumvented by coherently encoding photonic amplitude information into qubit registers and applying quantum algorithms to process the stored information from asynchronously arriving optical signals. 
As a specific example, we develop a quantum algorithm for imaging unresolved point sources and apply it to exoplanet detection. We demonstrate that orders-of-magnitude improvements in performance can be achieved under realistic imaging conditions using relatively small scale quantum processors.       
\end{abstract}

\maketitle

\section{Introduction \label{sec:intro}}
The ability to extract information from faint light sources is vital across a wide range of applications, ranging from  molecular imaging~\cite{Liu2015} to satellite surveillance~\cite{Choi2024}, and astrophotography~\cite{Bruce2006,Soummer2012,Carter2023,Mullally2024}. Enhancing key performance metrics, such as resolution and signal-to-noise ratio (SNR), is essential for improving image quality and expanding the scope of observable phenomena. Such improvements could deepen our understanding of molecular dynamics, enable the detection of distant astronomical objects, and advance scientific discovery more broadly. 

In conventional optical imaging, the system records classical measurement data by integrating over detected signals. Classical post-processing techniques, such as principal component analysis or phase retrieval algorithms, are then applied to extract relevant information, including direct images, spectral features, or wavefront information for adaptive optics~\cite{Bruce2006,Soummer2012,Carter2023,Mullally2024}. However, restricting the system to classical data integration followed by post-processing prevents direct interference of asynchronously arriving optical signals, imposing a fundamental limitation on the capabilities of current techniques. 

Recent studies have shown that learning about a physical system through post-detection classical analysis can be significantly less efficient than approaches that leverage quantum information processing. In particular, using a quantum memory to store multiple samples of a quantum state and applying quantum algorithms before detection have been shown to offer significant advantages in certain settings~\cite{Khabiboulline2019,Huang2022,Allen2025}. In light of the rapid progress in quantum computing hardware across various platforms~\cite{Seetharam2023,Bluvenstein2024,reichardt2024,acharya2024}, quantum-enhanced sensing is becoming an increasingly viable potential application of quantum technologies. 
However, identifying specific, practical sensing tasks where quantum processing could yield clear performance gains remains an outstanding challenge in quantum science and engineering.

\begin{figure}[t]
    \centering
    \includegraphics[width=\columnwidth,keepaspectratio]{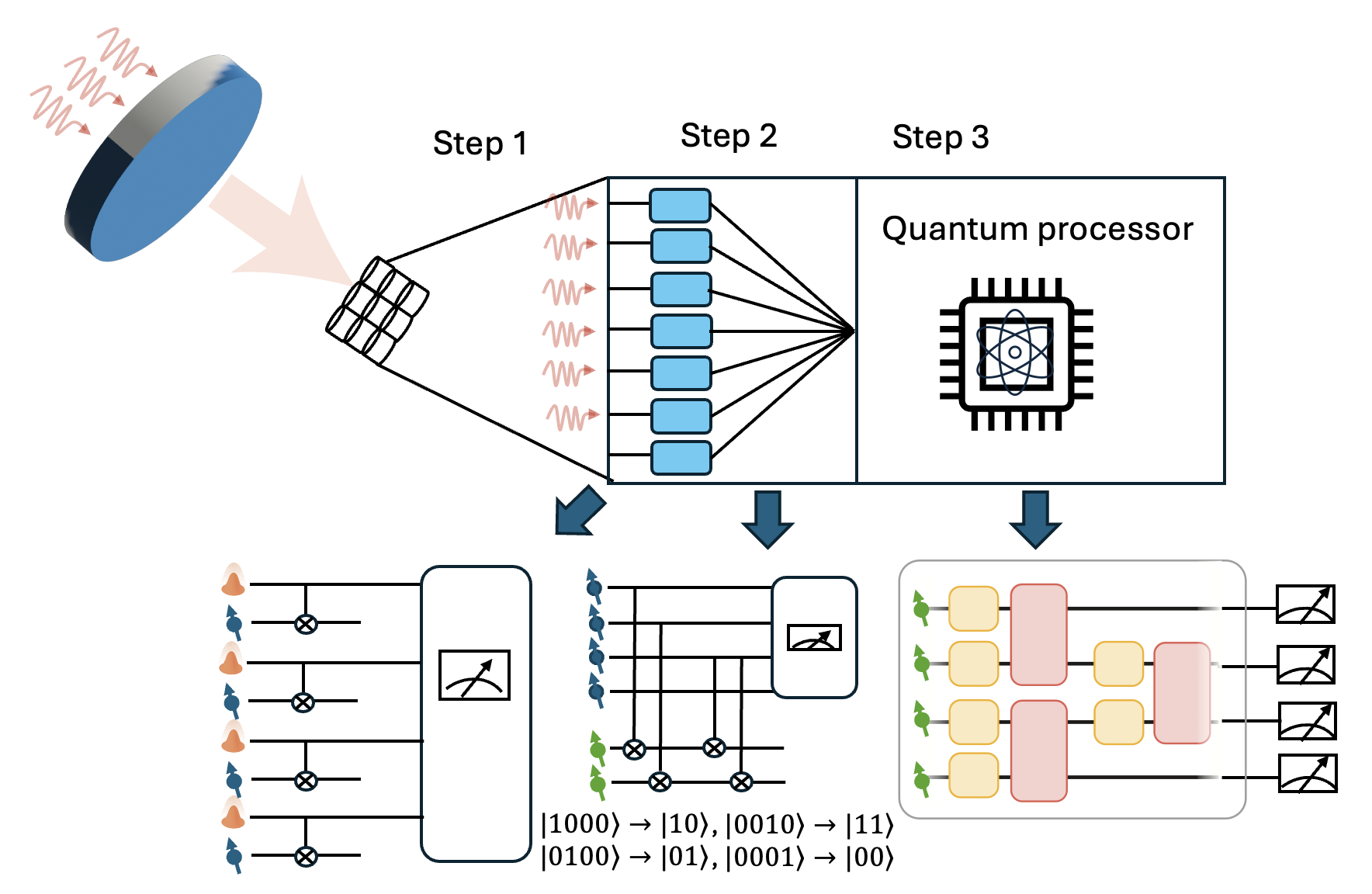}
    \caption{\justifying Sketch of a quantum processing enhanced imaging system. Step 1: The quantum state of the light collected through the optics is mapped to a pixel-qubit register in a heralded way by means of qubit-photon controlled gates followed by joint detection of the photonic modes. Step 2: For a weak optical signal where only a single photon is coherently distributed across the detection modes, the information can be compressed into a logarithmic number of memory qubits using a unary-to-binary encoding. Step 3: Quantum processing of the received light to extract the parameters of interest with higher SNR than possible from classical direct detection and post-detection processing.}
   \label{fig:figure0}
\end{figure}

In this work, we demonstrate that quantum information processing can offer fundamentally new and widely applicable advantages for imaging weak optical fields. Specifically, we show that encoding photonic amplitude information into a qubit register enables the use of quantum algorithms, analogous to classical post-detection methods in conventional imaging, to extract the desired signal features. Crucially, by performing quantum processing prior to measurements, we can avoid the use of tomographic methods and  accumulation of shot noise that limits the performance of classical techniques, which  leads to a significant improvement in SNR scaling with system dimensionality. As a specific example, we develop a novel quantum algorithm, combining techniques from quantum principle component analysis~\cite{Lloyd2014}, quantum signal processing~\cite{low2017,Motlagh2024}, and block encoding~\cite{low2019qubitiz, Martyn2021} for imaging of unresolved point sources and apply it to the practically important task of exoplanet imaging in astronomy. We show that, under typical imaging conditions, several orders of magnitude improvement in SNR can be achieved through the quantum processing with realistic resources. This enhancement translates to a corresponding reduction in signal integration time, which in turn relaxes telescope stability requirements and enables narrower bandwidths in science measurements allowing for more selective and efficient detections.

Our analysis reveals a promising application of quantum-computing enhanced sensing for weak field optical imaging. While here we focus on exoplanet detection as a specific application, the general techniques presented could be extended and applied to other tasks such as molecular imaging, satellite monitoring and adaptive optics where the increased SNR from quantum processing could lead to near-term demonstrations of practical quantum advantage.  

\section{Quantum advantage for optical imaging}

The key steps of quantum processor–enhanced optical imaging are illustrated in Fig.~\ref{fig:figure0}. In the first stage, a weak optical signal carrying the information of interest is collected and focused into a set of detection modes. In classical imaging with, e.g. a CCD camera, each mode would correspond to a pixel that records light intensity.

In contrast, in our approach the full amplitude information of each mode is first coherently mapped onto a  pixel-qubit register using qubit-photon entangling gates, as demonstrated in both atomic~\cite{Welte2018} and solid-state~\cite{Bhaskar2020,Knaut2024,wei2025} systems. In practice, this is achieved via photon reflection from a cavity coupled system containing a qubit (such as electronic or nuclear spin), that entangles the presence/absence of a photon in a mode with the corresponding qubit. The reflected signals are then coherently combined and measured in a joint basis chosen to preserve coherence across the modes by revealing only photon presence and ``erase'' the information about the mode origin of the photon.

In the second step, the collective quantum state of the pixel-qubit register is compressed into a smaller memory qubit register for efficient storage and processing. For weak signals, where typically a single photon is collected across all modes, the amplitude distribution over $D$ modes can be compressed into only $\log(D)$ qubits~\cite{Khabiboulline2019}. For instance, a 32 × 32 pixel image requires just 10 qubits. Note that in this approach no classical-to-quantum data encoding is needed, since the input light is already stored as a quantum state.

In the third step quantum algorithms are applied to process the stored amplitudes, enabling  unitary transformations prior to the final measurement. 
 This allows optical mode manipulations, such as those used in mode sorting~\cite{Tsang2019,Dutton2019,PRA2017,Bao2021} or adaptive optics~\cite{davis2012}, to be implemented as quantum circuits. 
A fundamental advantage of quantum processing emerges when storing multiple photons arriving over time. \jb{Unlike recent studies focused on uncontaminated signals such as resolving the angular separation between point sources \cite{Tsang2019} or estimating properties of the principal component of a quantum state \cite{Huang2022}, we address the practically relevant regime in which the signal of interest is embedded in a strong background noise.} Here, classical imaging typically relies on tomographic methods for extracting effective models describing the system, such as the Point Spread Function (PSF) or characterization of speckle noise for adaptive optics, mode sorting and differential imaging, these estimations require high SNR and impose a sampling complexity that scales with the system’s dimension, typically set by the number of pixels capturing the image. Moreover, post-detection processing accumulates the shot noise across multiple pixels which further degrades the SNR. 

Quantum processing offers a qualitatively different paradigm: extracting specific features from quantum states without estimating the full classical model of imaging target and noise. An illustrative example is the quantum SWAP test~\cite{barenco1997,Buhrman2001}, which can determine the similarity of two quantum states without the need for the full state tomography. Extensions of this idea to algorithms such as quantum principal component analysis~\cite{Lloyd2014} enables the ability to estimate observables from the principal components of a quantum state, without having to estimate the full structure of that state. It was demonstrated previously that this algorithm can lead to an improvement in SNR that scales with system dimensionality~\cite{Huang2022} since it circumvents the need of tomographic state reconstruction.  In our approach, 
we combine techniques from quantum principle component analysis~\cite{Lloyd2014} and quantum signal processing~\cite{low2017,Motlagh2024} to sort the incoming light of unresolved point sources into the eigenbasis of the PSF allowing for a direct measurement of observables in this basis without having to estimate the PSF basis structure.

\begin{figure*}[t]
    \centering
    \includegraphics[width=\textwidth,keepaspectratio]{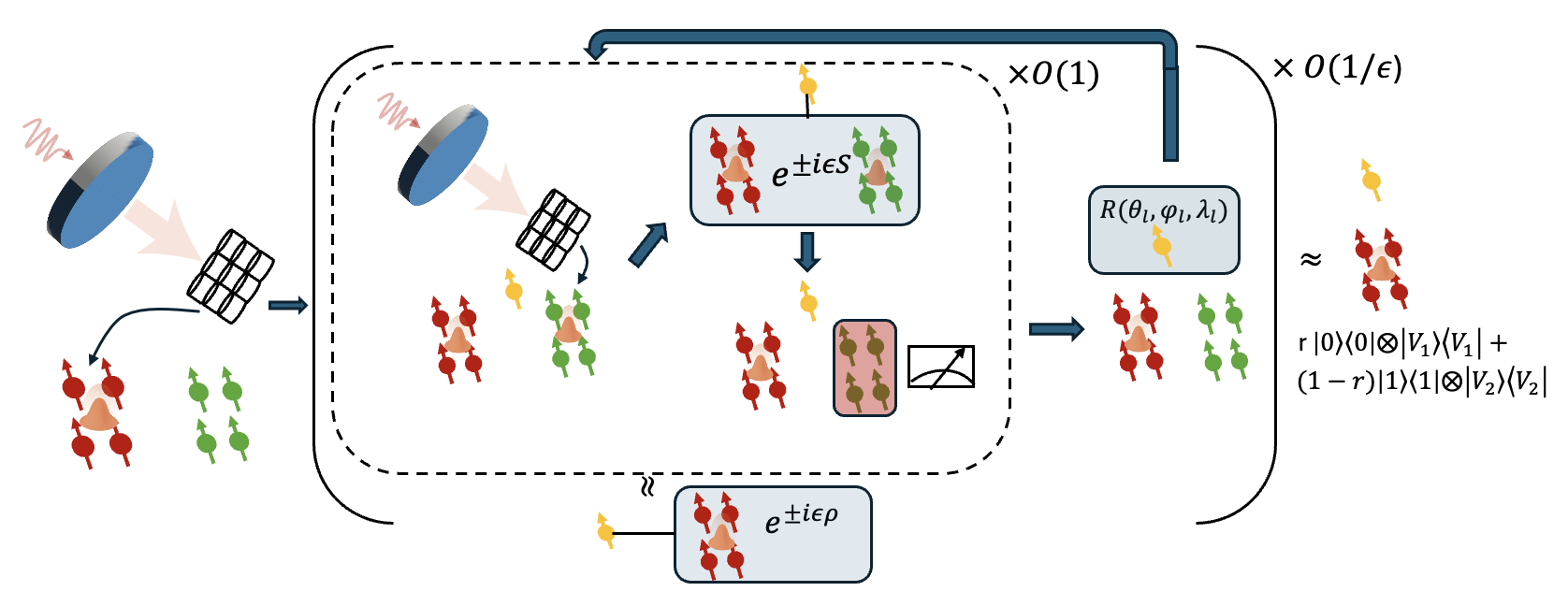}
    \caption{\justifying Quantum algorithm for sorting weak  light signal into the PSF eigenbasis. Left to right: the quantum state of an incoming photon is first stored in a compressed form in a qubit memory register (red qubits). The state of a subsequent photon is then stored in a compressed form in a second memory register (green qubits). An auxiliary qubit (yellow) mediates a controlled $e^{\pm i \epsilon \mathcal{S}}$ gate between the two registers (with $\mathcal{S}$ being a swap operator), after which the register holding the later photon is measured, re-initialized and then used to store a subsequent photon state. To sort eigenstates to a given precision $\epsilon$, this procedure is repeated $O(1)$ times times to approximate the controlled application of $e^{\pm i\epsilon\rho}$ to the first register. A single-qubit rotation is then applied to the auxiliary qubit, and the approximate application $e^{\pm i\epsilon\rho}$ is repeated. The entire sequence is executed $O(1/\epsilon)$ times (up to logarithmic factors; see main text) to sort the initial photonic state into the PSF eigenbasis with error bounded by $\epsilon$: $\mathrm{QSP}_{\exp(ix\rho),f_s} (\rho \otimes \ket{0}\bra{0}) \approx r\ket{V_1}\bra{V_1} \otimes \ket{0}\bra{0} + (1-r)\ket{V_2}\bra{V_2} \otimes \ket{1}\bra{1}$. Here, $r$ is related to the relative star-to-exoplanet intensity. The single qubit rotations and phase ($\pm$) of $e^{\pm i\epsilon\rho}$ are chosen to approximate the application of a step function such that a final measurement of the auxiliary qubit reveals which eigenstate the register holds. }
   \label{fig:processing_fig}
\end{figure*}

Specifically, we repeatedly apply controlled operations between a single stored photon and subsequently received photons to approximately implement a state-dependent unitary following the method of quantum principal component analysis~\cite{Lloyd2014}. This unitary is applied conditionally on an auxiliary qubit, which undergoes single-qubit rotations between the controlled operations. In this way, the state of the auxiliary qubit becomes correlated with the PSF eigencomponents of the photonic state. A final measurement of the auxiliary qubit then reveals which eigenstate the stored state is in without requiring any knowledge of the structure of the specific state (see Fig.~\ref{fig:processing_fig}).   
Combining this approach with standard Gram-Schmidt orthogonalization and block encoding~\cite{low2019qubitiz, Martyn2021} allows us to estimate observables of the individual sources directly. \jb{Similar to the quantum SWAP test, our approach avoids tomographic reconstruction and the associated accumulation of shot noise in post detection processing, which gives a quantum advantage that scales with the dimensionality of the system. Instead of estimating the noise structure from measurements, as required in classical background subtraction methods, our technique removes the background coherently at the quantum level. In classical methods, the estimation of the noise is itself limited by shot noise and this uncertainty propagates directly into any later processing that attempts to subtract the background from the signal. In contrast, our approach removes the need for this step, so that shot noise enters only through the final measurement of the observable of interest.}

The potential of this approach can be illustrated for a typical imaging scenario in which the star-exoplanet system is resolved on a 10×10 pixel array~\cite{Carter2023, Mullally2024}. Assuming that a coronagraph or starshade reduces the starlight at the detector to roughly ten times the brightness of the exoplanet, our analysis described below reveals that  achieving an SNR of 10 requires 3–4 orders of magnitude fewer detected photons when using quantum processing compared to conventional classical tomographic methods. Notably, this can be accomplished with an array of 100 pixel-qubits, a modest quantum processor consisting of around 36 memory qubits and total two-qubit gate counts on the order of hundreds, which is within the capabilities of current quantum information systems.   

\section{Quantum description of stellar light}
\label{sec:formulation}
We consider an illustrative example of imaging an unresolved system of a star with an accompanying exoplanet. As described in Refs.~\cite{nair2016interferometric}, the light received from the exoplanet and star in a certain detection time window is well approximated as an incoherent mixture of vacuum and single photon contributions of the form 
\begin{equation}\label{eq:tot_incoming_state}
\begin{split}
\rho_{\textrm{tot}} = (1 - \delta_{\textrm{vac}})\ket{\textrm{vac}}\bra{\textrm{vac}}+\\
\delta_{\textrm{vac}} \left(b \ket{\psi_1}\bra{\psi_1} + (1-b) \ket{\psi_2}\bra{\psi_2}\right),
\end{split}
\end{equation}
in the weak signal limit. Here $\ket{\psi_1}$ ($\ket{\psi_2}$) is the single photon state received from the star (exoplanet). The quantity $\delta_{\textrm{vac}}$ determines the probability of receiving a photon and $b$ is the relative brightness of the star and exoplanet. We note that while the star is in general much brighter than the exoplanet, the use of coronagraphs and starshades can suppress the starlight such that the relative brightness at the detector is comparable. We asssume the light to be  quasi-monochromatic due to the narrow bandwidth of the quantum pixels ($\sim$ 1 GHz for SiVs~\cite{Knaut2024,Bhaskar2020}), which imposes a spectral filtering of the light. 

Both light from the star and the exoplanet have a wavefunction ($\psi_k$) determined by the PSF of the telescope and the location of the respective objects on the sky. Since the distance between the star/exoplanet and the telescope is very large, we will ignore the finite size of the star and exoplanet and model the light as approximately plane waves with perfect spatial coherence, which is why the single photon components are modeled as pure states in Eq.~(\ref{eq:tot_incoming_state}). We provide more details on the PSF function and how this determines the wavefunction in Appendix~\ref{app:psf_model}. 

The states $\ket{\psi_1}$ ($\ket{\psi_2}$) can be represented in a basis consisting of the ``position" states $\ket{x,y} = a^{\dagger}_{x,y}\ket{\textrm{vac}}$, where $a^{\dagger}_{x,y}$ is a creation operator of the photon at the location $x,y$ of the detector plane and $\ket{\textrm{vac}}$ is the vacuum state. We assume that the incoming state is detected by a pixelated sensor, which projects the incoming state onto a $2$-dimensional $N \times N$ regular lattice of photonic spatial modes, $\phi_{m,n}(x,y)$ with $m,n\in\{1,\ldots,N\}$. We therefore express the incoming state in  pixelated basis states of the form:
\begin{equation}\label{eq:pixel_basis}
\ket{m,n} = \int_{\mathfrak{P}_{m,n}} \phi(x,y) \ket{x,y}  dxdy
\end{equation}
where $\mathfrak{P}_{m,n}$ denote the area of a single pixel and $\int_{\mathfrak{P}_{m,n}} |\phi(x,y)|^2 \, dxdy=1$.
An incoming photon with a wave function $\psi(u,v)$ induces the state of the detector:

\begin{equation}\label{eq:post_pixelation_state}
 \ket{\psi} = \frac{1}{\sqrt{\eta}}\sum\limits_{m,n = 0}^{N-1} \psi_{m,n} \ket{m,n},
\end{equation}

where the coefficients $\psi_{m,n}$ are given by

\begin{equation}\label{eq:pixelation_coefficients}
\psi_{m,n} = \int_{\mathfrak{P}_{m,n}} \phi^*(x,y)\psi(x,y) \, dxdy,  
\end{equation}

and $\eta=\sum|\psi_{m,n}|^2 $ is the total detection efficiency. 

\section{Mapping to a qubit register}

In classical optical imaging, a CCD camera records the light intensity per pixel corresponding to measuring the projections $\bra{m,n}\rho_{\text{tot}}\ket{m,n}$. Instead, we imagine to replace the pixels with quantum photon-spin transducers capable of mapping the incoming amplitude of the light into a pixel-qubit register. This can be implemented through optical reflection from a qubit-coupled optical cavity system as experimentally demonstrated with solid-state defect centers~\cite{Bhaskar2020,Stas2022,Knaut2024}. 

Specifically, for a properly tuned cavity-spin system, the reflected photonic amplitude from the cavity-qubit system can be tuned such there is a $\pi$ phase difference between reflection from qubit state $\ket{0}$ and $\ket{1}$~\cite{Welte2018}. Initializing the $m,n$-pixel qubit in state $\ket{0}$ and applying a single qubit Hadamard gate at the start and end of the detection window will ideally result in the transformation
\begin{eqnarray}
&&\left(\alpha\ket{\text{vac}_{m,n}}+\beta\ket{1_{m,n}}\right)\ket{0}_{m,n}\to \nonumber \\
&&\alpha\ket{\text{vac}_{m,n}}\ket{0}_{m,n}+\beta\ket{1_{m,n}}\ket{1}_{m,n},
\end{eqnarray}
for some arbitrary photonic state $\left(\alpha\ket{\text{vac}_{m,n}}+\beta\ket{1_{m,n}}\right)$. How well this mapping is implemented depends on the physical hardware and in particular, the strength of the qubit-cavity coupling. Recent experiments with Silicon Vacancy centers with have demonstrated similar qubit-photon operations with infidelities of only a few percent~\cite{Bhaskar2020,Stas2022,Knaut2024}.   

Performing the above transformation at all pixels would entangle the incoming photonic state with the pixel qubits. To complete the transfer of the amplitude information to the pixel qubits, the reflected light is measured in the Fourier basis $\{U_\text{QFT}\ket{m,n}\}$, where $U_{\text{QFT}}$ is the unitary matrix corresponding to the Quantum Fourier Transform. This will herald that a photon was received without revealing at which pixel it was received and can be implemented by interfering the reflected light from the pixel-qubits on a set of optical beam splitters and phase shifters before detection~\cite{reck1994experimental, barak2007quantum}. Consequently, the vacuum component of the state in Eq.~\eqref{eq:tot_incoming_state} can be discriminated and the pixel-qubits will be prepared in the state
\begin{equation}\label{eq:incoming_state}
\rho = b \ket{\psi_1}\bra{\psi_1} + (1-b) \ket{\psi_2}\bra{\psi_2}.
\end{equation} 
up to (known) single qubit phase corrections dictated by the photonic Fourier basis measurement outcome. 

After mapping of the light into the pixel-qubits, the state is transferred to a quantum processor. While this processor could, in principle, be implemented with the same hardware as the pixel-qubits, this may be suboptimal since the quantum pixels are optimized for receiving light and not for general quantum processing. As we describe below, the information from the pixel-qubits can in that case be transferred via quantum teleportation to another hardware for processing.      


\section{Quantum processing}\label{sec:QSP_for_hev}

The steps outlined above enable us to store multiple samples of the incoming state (Eq.~(\ref{eq:incoming_state})) in the quantum processor in a heralded manner. This opens new possibilities for separating the components from the star and the exoplanet by means of quantum processing. 

Quantum phase estimation (QPE)~\cite{kitaev1995} is a well-known quantum algorithm that allows for efficient estimation of eigenenergies of a given Hamiltonian, $H$. The algorithm operates by applying controlled unitaries of the form $e^{-iHt}$ to a given state of the system, corresponding to evolving the state for a time $t$ under $H$, conditioned on an auxiliary qubit register. As a result, the state is decomposed into its eigenstate components, and a final measurement of the auxiliary qubits projects the system onto one of these components while providing an estimate of the corresponding eigenenergy.

Quantum principal component analysis (QPCA)~\cite{Lloyd2014} utilizes QPE to decompose an unknown quantum state $\tilde{\rho}$ into its eigenvectors. Specifically, given access to controlled unitaries of the form $e^{-i\tilde{\rho} x}$, one can decompose an unknown state $\tilde{\rho}$, into its eigenstates and obtain estimates of the corresponding eigenvalues. Importantly, Ref.~\cite{Lloyd2014}, demonstrated that one can obtain arbitrarily precise approximations of $e^{-i\tilde{\rho} x}$ with no prior knowledge of $\tilde{\rho}$ provided access to a sufficient number of samples of the state.


We leverage this idea to sort the incoming light into its eigenbasis. Note that the eigenmodes of the PSF, $\ket{V_1},\ket{V_2}$ corresponds to the eigenmodes of $\rho$, i.e. 
\begin{equation}
\rho=r_1\ket{V_1}\bra{V_1}+r_2\ket{V_2}\bra{V_2},
\end{equation}
with eigenvalues $r_1$ and  $r_2=1-r_1$.

The procedure involves the ability to apply the conditional gate $e^{-ix\rho}$ where $x$ is a (small) real number. As shown in Ref.~\cite{Lloyd2014}, this is possible given access to a collection of states through the following approximation. Let $\mathcal{S}$ denote the usual SWAP gate between identical multidimensional spaces containing states $\rho$ and $\sigma$. For sufficiently large $k \in \mathbb{N}$:
\begin{equation}\label{eq:software_state_sim}
\begin{split}
&\mathrm{Tr}_{1} \left[ \exp(-i\mathcal{S}/k) \rho \otimes \sigma \exp(i\mathcal{S}/k) \right] = \sigma - i \frac{1}{k} [\rho ,\sigma] + O\left(\frac{1}{k^2}\right)\\
& = \exp(-i\rho/k) \sigma \exp(i\rho/k)+ O\left(\frac{1}{k^2}\right),
\end{split}
\end{equation}
where $\mathrm{Tr}_1$ denotes the trace over system 1. Applying this operation $n$ times (using $n$ samples of $\rho$) results in the unitary $\exp\left(-i n/k \rho\right) \sigma \exp\left(i n/k \rho\right)$ with a precision of $O(n/k^2)$. This provides a method for obtaining an approximation of $e^{ix\rho}$ for any real $x$ with arbitrary precision. Achieving an approximation with precision $\varepsilon_g$ requires $O(x^2/\varepsilon_g)$ samples of $\rho$, but they are used sequentially, which means that only a qubit register capable of storing two samples of $\rho$ is required in our case. Higher-order terms in Eq.~(\ref{eq:software_state_sim}) involve higher-order commutators suppressed by coefficients derived from the series expansion of the exponential function. Since these commutators involve only density matrices, the corresponding constants are on the order of $1$. Additionally, as stated in Refs.~\cite{Lloyd2014, kimmel2017}, converting this expression into a primitive gate count introduces a factor proportional to the logarithm of the dimension.  

The above approximation allows us to sort $\rho$ into its eigenbasis i.e. the PSF eigenbasis. This is achieved by substituting the $\mathcal{S}$ gate with a controlled $i\textrm{SWAP}x$ gate:
\begin{equation}\label{eq:iSWAPt}
U_{x}^{\mathcal{S}} = \ket{0}\bra{0} \otimes I  +  \ket{1}\bra{1} \otimes \exp(i\mathcal{S}x)
\end{equation}
and using an auxiliary qubit. 

In both QPE and QPCA, an auxiliary qubit register and a quantum Fourier transform are used to project onto the eigenstates and extract estimates of the corresponding eigenvalues. However, this is not necessary in our setting, where the goal is to prepare a collection of eigenstates that can be sampled directly. As we will show later, sampling from this collection allows us to obtain sufficiently accurate estimates of the eigenvalues. Our objective is therefore simply to distinguish the eigenstates from one another, which can be achieved more efficiently using quantum signal processing (QSP)~\cite{low2017,Motlagh2024}. 

The key idea of QSP is to encode a complex function of the eigenvalues of a controlled unitary into a quantum state by approximating the function with trigonometric polynomials. This encoding is implemented by repeatedly applying the controlled unitary, interspersed with single-qubit rotations on the control qubit, with the rotation angles determined by the coefficients of the approximating polynomial.  

In particular, consider an auxiliary qubit prepared in a state $\ket{q} = \alpha \ket{0} + \beta \ket{1}$. Let $U_{x}^{\mathcal{\rho}}$ denote controlled $e^{ix\rho}$. From Eq.~(\ref{eq:software_state_sim}), we obtain:
\begin{equation}\label{eq:cond_exp_rho}
\begin{split}
&\mathrm{Tr}_{2} \left[ U_{-x}^{\mathcal{S}} \left(\ket{q}\bra{q} \otimes \rho \otimes \sigma\right) U_{x}^{\mathcal{S}}\right] \approx\\
&|\alpha|^2\ket{0}\bra{0} \otimes \sigma + |\beta|^2 \ket{1}\bra{1}\exp(-i\rho x) \otimes \sigma\exp(i\rho x) + \\
& \alpha \beta^* \, \ket{0}\bra{1} \otimes \sigma\exp(i\rho x)+\beta \alpha^*\ket{1}\bra{0} \otimes \exp(-i\rho x)\sigma = \\
&U_{x}^{\mathcal{\rho}} \left(\ket{q}\bra{q} \otimes \sigma\right) U_{-x}^{\mathcal{\rho}},
\end{split}
\end{equation}
where we have used the similar approximation of
\begin{equation}\label{eq:software_state_sim1}
\begin{split}
&\mathrm{Tr}_{1} \left[ \exp(-i\mathcal{S}x) \rho \otimes \sigma \right] = \sigma - i x \rho\sigma + O(x^2) = \\
& \exp(-ix\rho) \sigma + O(x^2). 
\end{split}
\end{equation}

Note that if $\sigma=\ket{V_k}\bra{V_k}$ is an eigenvector of $\rho$ with the eigenvalue $r_k$, the transformation corresponds to 
\begin{equation}\label{eq:cond_exp_rho}
\ket{q}\ket{V_k} \mapsto \left(\alpha\ket{0} + \exp(i r_k x)\beta\ket{1}\right) \ket{V_k} +O(x^2)
\end{equation}
Such conditional gates can be used for the QSP procedure~\cite{low2017,Motlagh2024} to implement a more complex transformation. Define a general one qubit rotation

\begin{equation}
R(\theta, \phi, \lambda) =
\begin{bmatrix}
e^{i(\lambda+\phi)}\cos(\theta) & e^{i\phi}\sin(\theta) \\
e^{i\lambda}\sin(\theta) & -\cos(\theta)
\end{bmatrix}
\end{equation}

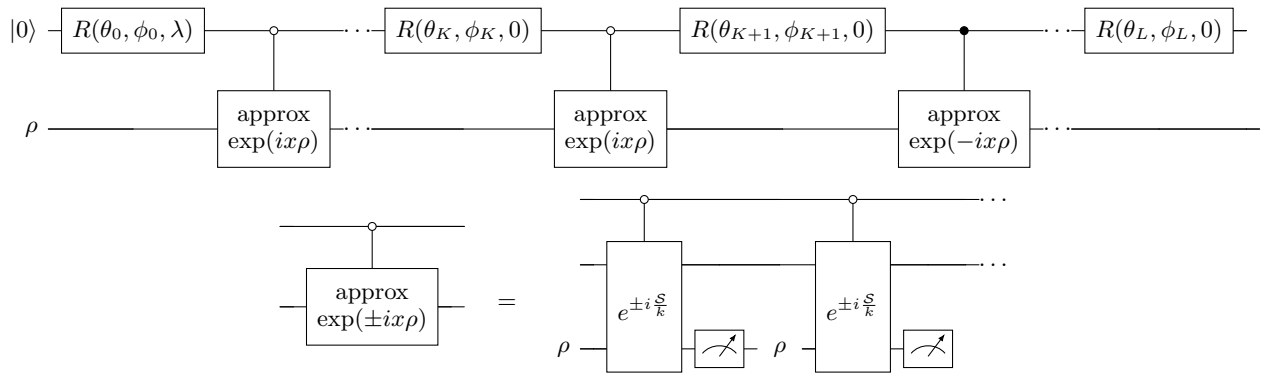
\begin{figure*}[t]         
  \centering                 

\begin{quantikz}[thin lines, scale=0.8, column sep=5pt]
    \lstick{$\ket{0}$} &
    \gate{R(\theta_0, \phi_0, \lambda)} & \octrl{1}  &\dots &
    \gate{R(\theta_K,\phi_K, 0)}  & \octrl{1} &
    \gate{R(\theta_{K+1},\phi_{K+1}, 0)} & \ctrl{1} &
     \dots & \gate{R(\theta_L,\phi_L, 0)} & \qw \\
    \lstick{$\rho$} &
    \qw & \gate[2]{\makecell{\text{approx}\\\exp(i x\rho)}} &\dots &
    \qw  & \gate[2]{\makecell{\text{approx}\\ \exp(i x\rho)}} &
     \qw & \gate[2]{\makecell{\text{approx}\\ \exp(-i x\rho)}} &
     \dots & \qw & \qw& \qw
\end{quantikz}


\vspace{5pt}

\begin{quantikz}[thin lines, scale=0.8, column sep=5pt]\label{circ:app_block}
    \lstick{} &  \qw & \octrl{1} &  \qw & \qw \\
    \lstick{} & \qw & \gate[2]{\makecell{\text{approx}\\ \exp(\pm i x\rho)}} & \qw &  \\
\end{quantikz}
$\quad=\quad$
\begin{quantikz}[thin lines, scale=0.8, column sep=5pt]\label{circ:app_block}
     \lstick{} &  \qw & \octrl{1} &  \qw & \qw  & \octrl{1} & \qw & \qw &\ldots  \\
    \lstick{} & \qw & \gate[2]{e^{\pm i \frac{\mathcal{S}}{k} }}  & \qw &\qw & \gate[2]{e^{\pm i\frac{\mathcal{S}}{k}}} & \qw & \qw & \ldots\\
    \lstick{$\rho$} & \qw & \qw  & \meter{} & \midstick[1]{$\rho$} & \qw & \meter{}  \\
\end{quantikz}
 \caption{\justifying (top) The QSP circuit requires two registers of the same dimension and one ancillary qubit. One register stores the first received state $\rho$, while the other stores subsequent states as they arrive at later times. (bottom) The approximation of the unitary $ e^{-i x\rho} $ is implemented in  $x\cdot k$ steps, to obtain a precision of $O(x/k^2)$. Each step applies a conditional $e^{-i\mathcal{S}/k}$ gate, followed by re-initializing the second register so that it is ready to receive another state. This operation implements a conditional $e^{-i \rho/k}$ gate to precision $O(1/k^2)$.  The trigonometric approximation of the Heaviside function involves both positive and negative powers of $e^{i x\rho}$. Following the QSP procedure suggested in Ref.~\cite{Motlagh2024}, the first $K\sim L/2$ controlled gates in the QSP circuit applies anti-controlled $e^{i x\rho}$ and the remaining gates applies controlled $e^{-i x\rho}$. As shown in the main text, achieving an approximation of the Heaviside function to precision $\epsilon$ requires $x\sim\epsilon$. Accordingly, we can choose $k\sim 1/\epsilon$ in our algorithm, implying that only $O(1)$ steps are needed to approximate $ e^{\pm i \epsilon \rho}$. However, since we need to implement $L\sim 1/\epsilon$ of these gates, the total number of such gates (and photons) required scales as $O(1/\epsilon)$. }
  \label{fig:QSP_circuit_fig}
\end{figure*}

Given a trigonometric polynomial $f$, such that $|f|\le 1$, it is possible to determine rotational angles $(\theta_0, \ldots,\theta_L), (\phi_0, \ldots,\phi_L)$, and $\lambda \in \mathbb{R}$, such that the operation sequence in Fig.~\ref{fig:QSP_circuit_fig} applies the following transformation to the eigenvectors $\ket{V_k}$ of unitary $\exp(ix\rho)$ with the eigenvalue $e^{ir_k x}$
\begin{equation}\label{eq:eigenv_QSP}
\begin{split}
\mathrm{QSP}_{\exp(ix\rho),f}(\ket{V_k}\otimes \ket{0}) \approx f(r_kx) \ket{V_k} \otimes \ket{0} +\\
g(r_kx)\ket{V_k} \otimes \ket{1},
\end{split}
\end{equation}
where $g(y)$ has the property $|g(y)|^2 + |f(y)|^2 = 1, \forall y$.

By taking $f$ to approximate the Heaviside step function, it is possible to filter eigenvectors with desirable phases. For $s \in (0, \pi)$ define a shifted periodic Heaviside step function as follows:
\begin{equation} \label{eq:pseudo_sign}
\Theta_s(\tau) = 
\begin{cases} 
0, & \text{if}\quad \tau \in (-\pi + s, s), \\ 
1, & \text{if}\quad \tau \in (s, \pi + s), \\ 
\frac{1}{2}, & \text{if}\quad \tau = s; -\pi + s; \pi + s,
\end{cases} 
\end{equation}
continued periodically over all real values of $\tau$. A sequence of trigonometric polynomials $f_L$ approximating the Heaviside function based on the Chebyshev expansion is considered in Ref.~\cite{Chuang2017}. It provides a near-optimal approximation rate. If the approximations are sufficiently accurate we get:
\begin{equation} \label{eq:filtered_rho}
\begin{split}
\mathrm{QSP}_{\exp(ix\rho),f_s} (\rho \otimes \ket{0}\bra{0}) \approx r\ket{V_1}\bra{V_1} \otimes \ket{0}\bra{0} +\\
(1-r)\ket{V_2}\bra{V_2} \otimes \ket{1}\bra{1},
\end{split}
\end{equation} 
where we defined $r=r_1$ to ease the notation. This allows for the sampling of $\ket{V_1}$ and $\ket{V_2}$ by measuring the auxiliary qubit.

\begin{figure}[h]
    \centering
    \begin{tikzpicture}[scale=2,>=stealth]

  \pgfmathsetmacro{\xMin}{0}           
  \pgfmathsetmacro{\xMax}{3.14159}    
  \pgfmathsetmacro{\yMin}{-0.3}       
  \pgfmathsetmacro{\yMax}{1.0}        
  \pgfmathsetmacro{\sValue}{1.5}      
  \pgfmathsetmacro{\DeltaValue}{0.5}  

  \pgfmathsetmacro{\xLeft}{\sValue - \DeltaValue}   
  \pgfmathsetmacro{\xRight}{\sValue + \DeltaValue}  

  \fill[gray!30] (\xLeft,\yMin) rectangle (\xRight,\yMax+0.3);

  \draw[->] (\xMin,0) -- (\xMax+0.3,0) node[above]{$\tau$};
  \draw[->] (0,\yMin) -- (0,\yMax+0.5) node[left] {$f_s(\tau)$};

  \draw (0,0.1) -- (0,-0.01);
  \node[left] at (0,0) {$0$};
  
  \draw[thick] (\xMax,0.05) -- (\xMax,-0.05);
  \node[xshift=0pt, yshift=-8pt] at (\xMax,0) {$\pi$};
  
  \draw[thick] (\xLeft,0.05) -- (\xLeft,-0.05);
  \node[xshift=1.75pt, yshift=-8pt] at (\xLeft,0) {$s-\Delta$};
  
  \draw[thick] (\xRight,0.05) -- (\xRight,-0.05);
  \node[xshift=1.72pt, yshift=-8pt] at (\xRight,0) {$s+\Delta$};
  
  \draw (-0.03,0) -- (0.03,0);
  
  \draw[thick] (-0.05,1) -- (0.05,1);
  \node[left] at (0,1) {$1$};
  
  \draw[dashed, black] (0.1,1) -- (\xMax+0.3,1);

  \draw[fill, purple] (0.0,0.04) circle (0pt)
  node[xshift=4pt, yshift=+5pt] {$\delta$};

  \draw[dashed, purple] (0.0,1 - 0.04) -- (1.9,1 - 0.04);
 \draw[thick, purple] (-0.05,1 - 0.04) -- (0.05,1 - 0.04);
\node[xshift=13pt, yshift=-5pt, purple] at (0.0,1-0.04) {$1-\delta$};
  
  \draw[thick, purple]
    plot[domain=\xMin:\xLeft] (\x, {0.04});

  \draw[thick, purple]
    plot[domain=\xLeft:\xRight, samples=50, smooth]
    (\x, {
      0.04 + 0.92 * (3 * ((\x - \xLeft)/(\xRight - \xLeft))^2 - 2 * ((\x - \xLeft)/(\xRight - \xLeft))^3)
    });

  \draw[thick, purple]
    plot[domain=\xRight:\xMax] (\x, {0.96});

  \draw[fill] (\sValue,0) circle (1pt) node[above] {$s$};

\draw[fill, blue] (2.1,0) circle (1pt) 
    node[xshift=0pt, yshift=6pt] {$r\cdot x$};

\draw[fill, blue] (0.9,0) circle (1pt) 
    node[xshift=-12pt, yshift=8pt] {$(1-r) x$};

\end{tikzpicture}
    \caption{\justifying Absolute value of the polynomial $f_s$ which approximates $\Theta_s$ and the  positions of the phases of $\exp(ix\rho)$ which are used to sort $\rho$ into its eigenvectors. The parameters $\Delta$ and $\delta$ determines the region where $\delta<f_s(\tau)<1-\delta$.  }
    \label{fig:phases_position}
\end{figure}
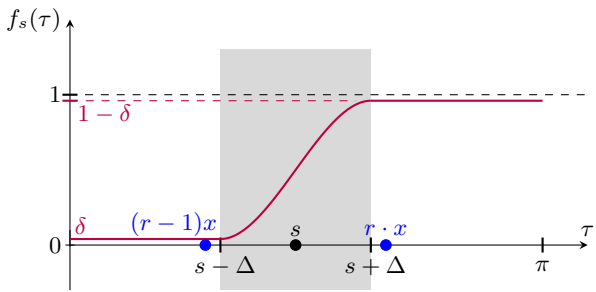

The complexity of such an approximation is determined by two parameters $0< \Delta < 1/2$ and $0 < \delta$ (see Fig.~\ref{fig:phases_position}). The parameter $\Delta$ sets the size of the  ``forbidden zone" , where we can't tolerate any of the eigenstates to lie within since the approximation of a step function is invalid; in this zone the continuous approximation function connects the step function discontinuity. The parameter $\delta$, sets how close the function value of the approximation is to unity/zero outside of the ``forbidden zone".  

In order to realize $\mathrm{QSP}_{\exp(ix\rho),f_s}$ with the approximating polynomial $|f_s(\tau) - \Theta_s(\tau)| \le \delta$ outside the ``forbidden zone" $(s-\Delta, s+\Delta)$, we need $O(\log(\delta^{-1})/ \Delta)$ conditional gates $\exp(ix\rho)$~\cite{Chuang2017, wang2024qpca}. The cost of such sampling in terms of samples of $\rho$ is determined by the relationship between $x$, which determines the number of gate cost, and $\Delta$, which affects the quality of the  QSP approximation. 

The forbidden zone must be entirely within the interval $(x(1-r), x r)$, so any $2 \Delta < 2 x r - x$ is suitable. Assuming that $r>3/4$ (typically $r\sim1$), we take $\Delta = x/4$ to obtain a complexity estimate in terms of $x$ and the precision, $\varepsilon_g$, of the approximation of the controlled gate $e^{ix\rho}$. We need $N_g = O(\log(\delta^{-1})/ \Delta) = O(\log(\delta^{-1})/ x)$ controllable gates, and for each gate we need $O(x^2/ \varepsilon_g)$ samples of $\rho$. Since we cannot use less than one state $\rho$, it means that for $x \propto \varepsilon_g^{\alpha_x}$ only $\alpha_x \le 1/2$ are allowed. Similarly, we cannot use less that one controllable gate and thus, only $\alpha_x\ge 0$ are allowed. In the range of meaningful exponents $0 \le \alpha_x \le 1/2$ the complexity is $O(\log(\delta^{-1})/ \varepsilon_g^{1-\alpha_x})$. Thus, the best possible choice is $x \propto \sqrt{\varepsilon_g}$ and the resulting complexity is $O(\log(\delta^{-1})/ \sqrt{\varepsilon_g})$. We achieve the precision $\epsilon$ of sampling $\ket{V_1}$ or $\ket{V_2}$ with a probability of $\delta$ to confuse which state ($\ket{V_1},\ket{V_2}$) is sampled, if the gate precision $\varepsilon_g $ satisfies
\begin{equation}\label{eq:main_eq_opt_compl}
\epsilon = N_g\cdot \varepsilon_g = O(\log(\delta^{-1})/ \sqrt{\varepsilon_g})\varepsilon_g,
\end{equation} 
and the total algorithm complexity is $O(\log(\delta^{-1})^2/ \epsilon)$.

We now have a procedure which allows us to sort the incoming light into the PSF eigenmodes. To achieve this to precision $\epsilon$, we store one received photon ($\rho$) and then use $O(\log(\delta^{-1})^2/\epsilon)$ consecutive photons to project into either $\ket{V_1}$ or $\ket{V_2}$. Note that we only need to store two photons at any point in time. We obtain $\ket{V_1}$ with probability $r$ and $\ket{V_2}$ with probability $1-r$. Hence, the interpretation of this sampling as a binary Bernoulli scheme with two outcomes gives us the estimate of $r$ as we continuously sample. 

\section{Measurement procedure} \label{sec:meas}
Our ultimate goal is to develop a measurement procedure that directly measures a specific observable $\mathrm{O}$ on the state emitted by the exoplanet and the state emitted by the star, i.e., $\bra{\psi_j} \mathrm{O} \ket{\psi_j}, \ j = 1,2$. This observable could e.g. correspond to the spatial image of the exoplanet as resolved on the pixel basis or the  intensity for a specific molecular line corresponding to a key biosignature~\cite{robinson2019earthexoplanet}. For now, we will assume that the relative intensity $b$ is known to good enough precision as a priory - an assumption adopted in related literature~\cite{nair2016interferometric,deshler2024achieving}. However, $b$ can also be estimated directly using quantum processing provided that we have some prior information about the relation between the star and the exoplanet as described in Appendix~\ref{app:mes_b}. Such prior information could e.g. come from spectral models of the star and exoplanet which predicts that the relative brightness of the two are expected to vary by a known factor at two different frequencies due to the absence/presence of different molecular lines~\cite{robinson2019earthexoplanet}.     

Our procedure requires $b \neq 1$  and $r\neq 1/2$.  The first condition indicates the presence of two distinct sources, an exoplanet detection problem, and can be verified using the SWAP test (see Appendix~\ref{app:SWAP_test} for a discussion). The value of $b$, together with the measured $r$, determines the vectors $\ket{\psi_j}$ up to a relative phase factor, which is inaccessible in the mixed state. Nonetheless, we can fix the representations $\ket{\psi_1}$ and $\ket{\psi_2}$ of the classes in the projective space representing quantum states, such that $\arg\left(\braket{\psi_1}{\psi_2}\right) = 0$, hence $\braket{\psi_1}{\psi_2} = h> 0$. Obtaining an expression for the eigenvectors $\ket{V_j}$ is straightforward using standard Gram-Schmidt diagonalization. First, we construct an the orthonormal basis $\{\ket{Y_1},\ket{Y_2}\}$ from the vectors $\ket{\psi_j}$:
\begin{equation} \label{eq:Ys}
\begin{split}
\ket{Y_1} =& \ket{\psi_1};\\
\ket{Y_2} =&\frac{1}{\sqrt{a}}\left(\ket{\psi_2} - \braket{\psi_1}{\psi_2} \ket{\psi_1}\right) =\\
&\frac{1}{\sqrt{a}}\left( \ket{\psi_2} - h \ket{\psi_1}\right),
\end{split}
\end{equation} 
where $a = 1 - |h|^2 $. The expression for $\rho$ in this basis is
\begin{equation} \label{eq:rhoinYs}
 \rho = \begin{pmatrix}
(b + (1-b)|h|^2) & h\sqrt{a}(1-b) \\
h\sqrt{a}(1-b) & (1-b)a
\end{pmatrix}.
\end{equation}
We can now find expressions for the eigenvalues ($r$ and $1-r$) and eigenvectors of $\rho$ by diagonalizing the matrix in Eq.~(\ref{eq:rhoinYs}). The determinant of the matrix is $D = r(1-r)$, which gives us the following relation:
\begin{equation} \label{eq:determ_rel}
h^2 = \frac{D}{b(b-1)} + 1.
\end{equation}
There is a unique pair of eigenvectors of $\rho$ which have real coefficients in the basis of $\ket{\psi_j}$. They have the following expressions ($r_1=r, r_2=1-r$)
\begin{equation} \label{eq:v_k_eigenv}
\begin{split}
\ket{V_k} = \frac{\left( r_k - (1-b) \right)\ket{\psi_1} + h (1-b)\ket{\psi_2}}{\sqrt{{{\left(r_k +a\,{\left(b-1\right)}\right)}}^2 +a\,{h}^2 \,{{\left(b-1\right)}}^2}}=\\
\tilde{c}_{1,k}\ket{\psi_1}+ \tilde{c}_{2,k}\ket{\psi_2}.
\end{split}
\end{equation}
Denote the inverse coefficients for $\ket{\psi_j}$ in the basis of $\ket{V_k}$ as $c_{j,k}$. The normalization coefficient is given by:
\begin{equation} \label{eq:norm_coeffVK}
\mathcal{N}_k=\left({{\left(r_k +a\,{\left(b-1\right)}\right)}}^2 +a\,h^2 \,{{\left(b-1\right)}}^2\right)^{-\frac{1}{2}}.
\end{equation} 

Since the coefficients $c_{j,k}$ are real, the expectation value of any observable $\mathrm{O}$ has the following expression:
\begin{equation} \label{eq:mes_O}
\begin{split}
\bra{\psi_k}\mathrm{O}\ket{\psi_k} = |c_{1,k}|^2 \bra{V_1}\mathrm{O}\ket{V_1} +\\
|c_{2,k}|^2\bra{V_2}\mathrm{O}\ket{V_2} + 2 c_{1,k}c_{2,k} \Re{\bra{V_1}\mathrm{O}\ket{V_2}}.
\end{split}
\end{equation} 

From Eq.~(\ref{eq:mes_O}), we see that as long as we have good enough estimates of the coefficients $\{c_{1,k},c_{2,k}\}$ and are able to sample from the PSF eigenbasis, we can estimate any observable of either of the two sources. Importantly, we do not need detailed information about the full structure of the PSF eigenstates. This is precisely where the quantum advantage arises.

The meaning of the argument of the off-diagonal term $\bra{V_1} \mathrm{O} \ket{V_2}$ for the observable $\mathrm{O}$ is ambiguous, but it is fixed when measuring the observable in the superposition of the states $\ket{V_k}$ with a fixed relative phase, as in Eq.~(\ref{eq:mes_O}). All overlaps on the right-hand side (RHS) of Eq.~(\ref{eq:mes_O}) can be measured by sampling the states $\ket{V_k}$ and applying appropriate processing. In Appendix~\ref{app:mes_OBS}, we describe how to measure the off-diagonal term of an arbitrary observable using block-encoding~\cite{low2019qubitiz}. Block-encoding allows us to implement a non-unitary transformation by embedding it in a larger unitary transformation followed by post selection to obtain the off-diagonal terms.

In Appendix~\ref{app:mes_O_from_ref}, we demonstrate a scheme to measure $\bra{\psi_k}\mathrm{O}\ket{\psi_k}$ without sampling $\ket{V_2}$, provided that we know all overlaps on the RHS of Eq.~(\ref{eq:mes_O}) for a reference observable $\mathrm{O}_{\text{ref}}$. This approach has two major advantages. The measurement scheme in Appendix~\ref{app:mes_OBS} is a probabilistic procedure, so it needs to be applied only once to a well-chosen reference observable $\mathrm{O}_{\text{ref}}$, which can be selected based on an accurate telescope model to optimize the success probability. The second advantage is that the measurement scheme requires only a single cycle of expensive sampling (assuming $r\sim1$) of $\ket{V_2}$. The scheme is based on a generalization of the SWAP test circuit (\ref{eq:SWAP_test_state}) and consists of the following steps:

\begin{enumerate}
    \item Measure $\bra{V_k} \mathrm{O}_{\text{ref}} \ket{V_l}, \ k,l = 1,\, 2$ by sampling the states $\ket{V_k}$ and using block encoding to obtain the cross terms (see Appendix~~\ref{app:mes_OBS})
    \item Estimate the eigenvalue  $r$ from the rate at which $\ket{V_1}$ is obtained. From Eq.~(\ref{eq:determ_rel}), we determine $h$ and the coefficients $c_{j,k}$ in (\ref{eq:v_k_eigenv}).
    \item Measure the value of $\bra{V_1}\mathrm{O} \ket{V_1}$.
    \item The SWAP test $SW_{i}(\ket{V_1},\rho)$ allows to estimate $\bra{V_2}\mathrm{O} \ket{V_2}$.
    \item Perform the SWAP tests $SW_{1}(\rho,\rho)$ to obtain a system of equations that allows the determination of $\bra{V_1}\mathrm{O} \ket{V_2}, \ k = 1,\, 2$.
\end{enumerate}

Note that only step 1 requires direct sampling of $\ket{V_2}$; all other measurements can be performed using the available surplus of $\ket{V_1}$ states.

\section{Noise}\label{sec:Noise}

The form of the photonic state in Eq.~(\ref{eq:tot_incoming_state}) is arguably a simplistic model since it does not account for additional noise, which would typically be present in practice. Such noise can originate from stray photons from the environment and random thermal excitations in the single photon detectors. We will now show how the algorithm also allows to sort the signal from a general noise floor, making it compatible with realistic noisy detector.

Precise description of the noise is challenging and greatly depends on the physical setup. However, to illustrate possible noise effects, we consider a noise model, where for all pixel modes, the contribution of the noise is equal. Thus, our signal state $\rho$ is shifted with the maximally mixed state $\frac{1}{N^2} I$, where $I$ is an identity operator, proportional to some noise level $\gamma \geq 0$:

\begin{equation} \label{eq:noise_model}
\begin{split}
\tilde{\rho} = (1-\gamma) \rho + \frac{\gamma}{N^2} I
= [(1-\gamma) r + \frac{\gamma}{N^2}] \ket{V_1} \bra{V_1} + \\
[(1-\gamma) (1-r) + \frac{\gamma}{N^2}] \ket{V_2} \bra{V_2} + 
\sum_{k=3}^{N^2} \frac{\gamma}{N^2} \ket{V_k} \bra{V_k},
\end{split}
\end{equation}

where $\ket{V_k}, k \geq 3$ are the noise terms orthogonal to our signal. This model is also compatible with a depolarizing noise on the SiVs spins from e.g. limited coherence and faulty gate operations, restricted to the single excitation subspace (higher order excitations could in principle be discriminated through parity measurements on the spins).  

We assume that the noise is well characterized so that we have a good estimate of $\gamma$. Depending on the expected relative magnitudes of $\gamma$ and $r$, we can choose different approaches for sampling the eigenvectors. 

To filter the noise, we can apply the QSP techniques previously described multiple times to perform a binary search. The efficiency of the filtering depends on the spectral gaps between the eigenvalues. The smallest gap is expected between the second eigenvalue, $(1-\gamma) (1-r) + \frac{\gamma}{N^2}$ and the noise terms. 

Let $p_0 = [(1-\gamma) r + \frac{\gamma}{N^2}]$ and 
\begin{equation} \label{eq:rho_2}
\begin{split}
\tilde{\rho}_2 = \frac{1}{1-p_0} \bigg(\big[(1-\gamma) (1-r) + \frac{\gamma}{N^2}\big] \ket{V_2}\bra{V_2}+\\ \sum_{k=3}^{N^2} \frac{\gamma}{N^2} \ket{V_k}\bra{V_k}\bigg) = \tilde{p}_{2} \ket{V_2}\bra{V_2} +   \sum_{k=3}^{N^2} \tilde{p}_{\textrm{res}} \ket{V_k}\bra{V_k},
\end{split}
\end{equation}
such that $\tilde{\rho}=p_0\ket{V_1}\bra{V_1}+(1-p_0)\tilde{\rho}_2$. Performing the QSP procedure realizes the approximation of the following operation
\begin{equation} \label{eq:noisy_QSP}
\begin{split}
\tilde{\rho} \xrightarrow{\mathrm{QSP}_{\tilde{\rho}}} 
p_0 \ket{0}\bra{0} \otimes \ket{V_1}\bra{V_1} + 
(1-p_0)\ket{1}\bra{1} \otimes \tilde{\rho}_2,
\end{split}
\end{equation}

If the gap $(1-\gamma)(1 - r)$ is not prohibitively small, we can first obtain $\tilde{\rho}_2$ by measuring the ancilla in the state $\ket{1}$, and then apply the QSP procedure based on $\tilde{\rho}$ to $\tilde{\rho}_2$. This allows us to separate $\ket{V_2}$ from the noise, provided the spectral gap is also sufficiently large. The sampling complexity of this approach, to filter the eigenstates up to precision $\epsilon$, is (see Appendix~\ref{app:Noisy_compls} for details)
\begin{equation} \label{eq:2seqQSPcomplmain}
O\left(\frac{\left[1+(1-\gamma)^{-2} (1-r)^{-2} \right]\log(\delta)^2}{\epsilon}\right).
\end{equation}

Alternatively, we can apply the QSP procedure directly to $\tilde{\rho}_2$ by approximating the unitary $e^{ix\tilde{\rho}_2}$. Reweighting in equation~\eqref{eq:rho_2} with a factor of $\frac{1}{1 - p_0}$ effectively increases the spectral gap associated with the eigenvector $\ket{V_2}$ in this state. This scheme is suitable for low noise $\gamma \lessapprox (1 - r)$ and has a sampling complexity of
\begin{equation}
O\left(\frac{\log(\delta)^4 \left( 1 - r + \gamma r  - \frac{\gamma}{N^2}\right)^2}{\epsilon^2  \left(\left(2r-1 \right)   (1-\gamma)(1-r)\right)^2}\right)
\end{equation}
i.e. quadratically worse in $\epsilon$, but with a suppression of the factor $\gamma$.

\section{Sampling complexity}
 
We will now provide estimates of the overall sampling complexity  for estimating a general observable of the exoplanet, $\bra{\psi_2}\text{O}\ket{\psi_2}$ using our algorithm and compare it to standard tomographic techniques. The sampling complexity is defined as the total number of photons required (i.e. number of samples of $\rho$) to obtain an estimate up to a certain statistical error, $\varepsilon_{\textrm{st}}$. 

The sorting of $\rho$ into the eigenvectors produces imperfect samples of $\ket{V_k}$; denote these samples as $|\tilde{V}_k\rangle$ and denote the sampling error as $\epsilon = \| \ket{V_k} - |\tilde{V}_k\rangle \|$. If we obtain $M$ imperfect samples of $|\tilde{V}_{k,m}\rangle, \, 1\le m \le M$ of the state $\ket{V_k}$ we obtain a standard estimator of the mean
\begin{equation}\label{eq:stat_error_QSP_k}
\begin{split}
\langle\mathrm{O} \rangle_{k} = \frac{1}{M} \sum_{m = 1}^{M} \langle\tilde{V}_{k,m}|\mathrm{O} |\tilde{V}_{k,m}\rangle =\\
\frac{1}{M} \sum_{m = 1}^{M} \bra{V_{k,m}} \mathrm{O} \ket{V_{k,m}}  + \frac{1}{M} \sum_{k = 1}^{M} \iota_{k}(\delta + \epsilon) =\\
\bra{V_k} \mathrm{O} \ket{V_k} + \varsigma_{\textrm{st}} + (\delta + \epsilon) \mathbb{E}(\iota).
\end{split}
\end{equation}
Here, we have introduced the bounded random variable $\iota$ (with samples $\iota_{k}$ and mean $\mathbb{E}(\iota)$ ) to capture the statistical noise from sampling from the erroneous distributions introduced through the state preparation errors ($\delta$ and $\epsilon$). The standard estimator of the mean requires $M_{st} = O(\varepsilon_\textrm{st}^{-2})$ measurement records to achieve a statistical error $\varsigma_{\textrm{st}} \approx \varepsilon_\textrm{st} \textrm{Var}(\mathrm{O})$ in the estimation of the mean. We lack sufficient information on the statistical properties of $\iota$, excluding that it is bounded by some modest constant, which is non-trivial to estimate theoretically. Consequently, we need to keep the magnitude $\delta + \epsilon$ below $\varepsilon_\textrm{st}$, which determines the number of required $\rho$ states as $M_{\textrm{prep}} = O\left( \log(\varepsilon_{\textrm{st}}^{-1})^2\varepsilon_\textrm{st}^{-1}\right)$ samples. 

Since we assume $r\sim1$, the sampling complexity will be determined by the required samples of $\ket{V_2}$. In particular, if we sample $M$ states of $\ket{V_2}$ to estimate $\bra{V_{2}} \mathrm{O}_{\text{ref}} \ket{V_{2}}$ to a certain precision, the total excess of $\ket{V_1}$ states obtained is approximately $M/(1-r)$. This ensures that a sufficient excess of $\ket{V_1}$ states for the SWAP test measurements of the off-diagonal terms and direct measurements of $\bra{V_k} \mathrm{O} \ket{V_k}$ in Eq.~(\ref{eq:mes_O}) (see Section~\ref{sec:meas}). It is important to note that the $\ket{V_1}$ state can be easily separated from noise using the QSP procedure, so additional samples of this state does not increase the overall complexity, even in the presence of noise.  

Estimating $\bra{V_{2}} \mathrm{O}_{\text{ref}} \ket{V_{2}}$ to statistical error $\varepsilon_{\textrm{st}}$, requires $M_{st}=O(\varepsilon_{\textrm{st}}^{-2})$ samples of $\ket{V_2}$, each of which on average requires $M_{\textrm{prep}}/(1-r)$ samples of $\rho$. This provides us with a total sampling complexity of 
\begin{equation}\label{eq:qsp_sample_compl}
M_{QSP} = M_{st}M_{\textrm{prep}}/(1-r)=O\left( \frac{\log(\varepsilon_\textrm{st})^2}{ (1-r)\varepsilon_\textrm{st}^{3}}\right).
\end{equation}
to estimate a general observable $\bra{\psi_2}\text{O}\ket{\psi_2}$ to statistical error $\varepsilon_\textrm{st}$. Since we effectively utilize the redundancy in $\ket{V_1}$, the constant in this $O$ asymptotic estimate is determined by the complexity constant of the QSP procedure, which is close to unity~\cite{Motlagh2024}, and by the corresponding constant from the approximation of the controlled $e^{ix\rho}$, which, as stated above, is also of order unity. Other constants, such as the variance of the observable, would similarly affect all measurement schemes. In the presence of separable noise the complexity according to Eq.~(\ref{eq:2seqQSPcomplmain}) acquires an additional factor $\left[1+(1-\gamma)^{-2} (1-r)^{-2} \right]$.

We compare the sampling complexity of our method to standard tomography underlying techniques such as differential imaging~\cite{Soummer2012,Carter2023,Mullally2024,acharya2024} and methods that utilize the sorting of spatial modes of light based on prior information about the PSF~\cite{Tsang2019,Dutton2019,PRA2017}. Such methods require learning the classical description of the PSF and use this to isolate the contribution from the exoplanet. Thus,  methods of this class will require performing quantum state tomography on the density matrix $\rho$.  The isolation of the contributions from the individual sources can then be obtained by diagonalization to obtain a classical description of the state of the exoplanet and the star. 


Unlike our algorithm, quantum state tomography requires a sampling complexity that scales with the dimension of the quantum state. This is because tomography learns the classical description of the quantum state, i.e., the form of the eigenstates, while our algorithm samples from the eigenstates without learning their detailed form. In particular, if the image is resolved on a $N$x$N$ pixel array, the dimension of the single photon state $\rho$ is $N^2$. Performing quantum state tomography to a precision of $\varepsilon_{\mathrm{tom}}$ on a mixed state, $\rho$ of rank 2 (described by only two eigenstates) requires at least $O\left(4\log(\varepsilon_{\mathrm{tom}}^{-1}) \frac{N^2}{\varepsilon_{\mathrm{tom}}^2}\right)$ samples of $\rho$~\cite{Haah2017}, when restricting to detection of each $\rho$ individually (i.e. direct detection). In the case of additional realistic noise of the form in Eq.~(\ref{eq:noise_model}) the state has full rank of $N^2$  and tomography on the general state requires $O\left(\frac{N^6}{\varepsilon_{\mathrm{tom}}^2}\right)$ states~\cite{Haah2017}. 

Tomography provides us with a noisy classical description $\overline{\rho}$ of the true state, such that $||\overline{\rho} - \rho||_{*} \le \varepsilon_{\mathrm{tom}}$ where $||\cdot||$ is the trace norm. 
According to Weyl's theorem, this yields the same error scaling in the eigenvalues: $\left|r_{k} - r_{k}(\overline{\rho})\right| = O\left(\varepsilon_{\mathrm{tom}}\right)$. A good estimate for the deviation of eigenvectors is provided by the Davis-Kahan theorem \cite{Davis1970, YuWang_useful_DK2015}, for the eigenvector $\ket{V_{k}}$ of $\rho$ with the eigenvalue $r_k$ such an estimate is given in terms of the spectral gap $\mathrm{g}(r_k) = \min\limits_{r\in \mathrm{Spec}{\rho}}|r_k - r_j|$. For eigenvalues $r$ and  $1-r$, we obtain the estimate  
\begin{equation}\label{eq:davis_Kahan_est}
\left|\left|\,\ket{V_{k}} - \ket{V_{k}(\overline{\rho})}\right|\right| \le \frac{C\varepsilon_\mathrm{tom}}{(1-r)},
\end{equation}
for some positive constant $C$, where $\ket{V_{k}(\overline{\rho})}$ are the estimated eigenvectors. This estimate is known to be exact. In Appendix~\ref{app:Davis_Kahan}, we investigate the tightness of this bound numerically and show that equality is achieved with the constant $C$ being significantly greater than zero. From this we conclude that this upper bound provides a good estimate for the complexity scaling. 
 
Using the estimated eigenvectors and eigenvalues, one can estimate any observable of the exoplanet from the estimated state $\ket{\psi_2(\rho)} = c_{1} \ket{V_{1}(\overline{\rho})} + c_{2}\ket{V_{2}(\overline{\rho})}$. The statistical error of the estimate is
\begin{equation}\label{eq:stat_error_tom}
\overline{\mathrm{O}}_{\textrm{tom}} = \bra{\psi_2(\rho)}\mathrm{O}\ket{\psi_2(\rho)}=
 \bra{\psi_2}\mathrm{O}\ket{\psi_2} + O\left(\frac{ \varepsilon_{\mathrm{tom}} } {1-r} \right).
\end{equation}
Thus, for an estimate with precision $\varepsilon_{\mathrm{st}} \propto \frac{\varepsilon_{\mathrm{tom}}}{1-r}$, the total sampling complexity is  

\begin{equation}\label{eq:stat_error_tom}  
M_{\mathrm{tom}} = O\left(\frac{4 N^2 \log(\varepsilon_{\mathrm{st}}^{-1} (1-r) ^{-1})}{\varepsilon_{\mathrm{st}}^2 (1-r)^2} \right)  
\end{equation}  

Notably, tomography exhibits a better scaling in $\varepsilon_{\mathrm{st}}$ than the quantum algorithm but worse scaling in $(1-r)$ and $N$. Importantly, for realistic observations $\varepsilon_{st}\sim0.1$ due the low flux of photons while $N \sim 10$ and $r\sim 1$~\cite{Carter2023,Mullally2024}, which results in orders of magnitude reduction in sampling complexity using quantum processing.   
\begin{figure}[ht]
    \centering
    \includegraphics[width=\columnwidth,keepaspectratio]{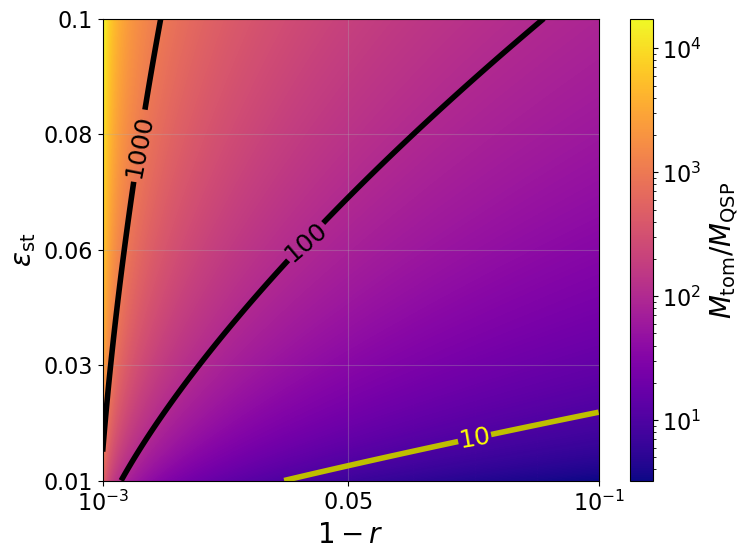}
    \caption{\justifying Ratio of sample complexities between the tomography and the QSP-based scheme in the absence of noise}
    \label{fig:rel_adv_N_sampls}
\end{figure}
\begin{figure}[ht]
    \centering
    \includegraphics[width=\columnwidth,keepaspectratio]{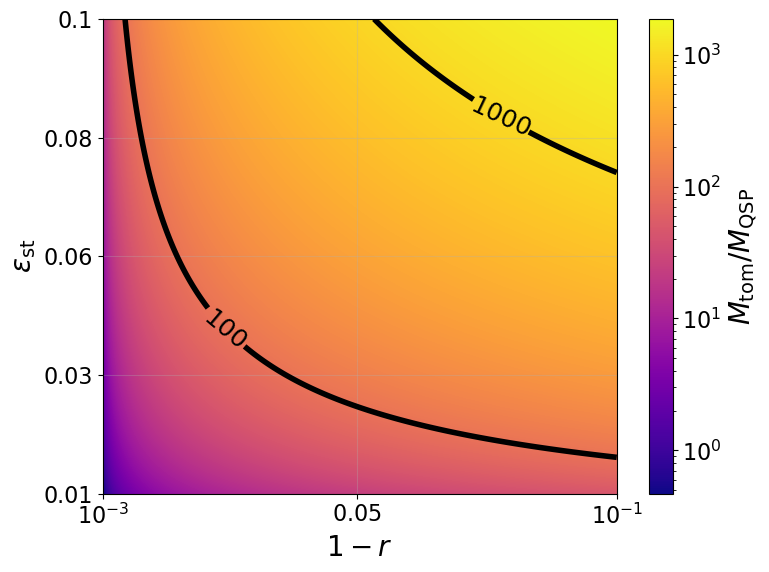}
    \caption{\justifying Ratio of sample complexities between the tomography and the QSP-based scheme in the presence of noise. The noise level is set to $\gamma = 10^{-3}$.}
    \label{fig:rel_adv_N_sampls2}
\end{figure}
In Figure~\ref{fig:rel_adv_N_sampls}, we plot the ratio $M_{\mathrm{tom}}/M_{\mathrm{QSP}}$ against $\varepsilon_{\mathrm{st}}$ and $r$ for a small $10 \times 10$ pixel array in the absence of detector noise. We see that quantum processing reduces the sampling complexity significantly over a broad range of $\varepsilon_{\mathrm{st}}$ and $r$.

In the realistic setting of noisy detection, the sampling complexity is substantially higher and depends on the rank of the noisy state. For the noise model considered in the Sec.~\ref{sec:Noise}, we obtain the following estimate for quantum state tomography:
\begin{equation}\label{eq:stat_error_tom_noisy}  
M_{\mathrm{tom}}^{\mathrm{full}} = O\left(\frac{N^6}{\varepsilon_{\mathrm{st}}^2 (1-r)^2} \right).
\end{equation}  
Furthermore, the sampling complexity of the QSP scheme changes from being $\propto  (1-r)^{-1}$ to a scaling of $(1-r)^{-3}$, i.e. slightly worse than the classical methods. The additional factor of $N^6$ in the complexity of tomography, however, means that even in the presence of noisy detection, quantum processing can provide several orders of magnitude reduction of the sampling complexity as seen from Figure~\ref{fig:rel_adv_N_sampls2}. We summarize the sampling complexity of both the classical tomographic methods and our algorithm in the presence/absence of detection noise in Tab.~\ref{tab:tab_complexity}. 

\begin{table}
\begin{center}
\begin{tabular}{ |c |c |c| } \hline
 Method & noise-free detection & noisy detection \\ \hline
 Classical & $O\left(\frac{4 N^2 \log(\varepsilon_{\mathrm{st}}^{-1} (1-r) ^{-1})}{\varepsilon_{\mathrm{st}}^2 (1-r)^2} \right)$ & $O\left(\frac{N^6}{\varepsilon_{\mathrm{st}}^2 (1-r)^2} \right)$ \\  \hline
 Quantum & $O\left( \frac{\log(\varepsilon_\textrm{st})^2}{ (1-r)\varepsilon_\textrm{st}^{3}}\right)$ & $O\left( \frac{\log(\varepsilon_\textrm{st})^2}{ (1-\gamma)^2(1-r)^3\varepsilon_\textrm{st}^{3}}\right)$   \\ \hline
\end{tabular}
\end{center}
 \caption{\justifying Sampling complexity of classical tomographic methods (Classical) and our quantum algorithm (Quantum) under noise-free and noisy detection. Sampling complexity is defined as the number of photons required to estimate an observable of the exoplanet within a statistical error $\varepsilon_\textrm{st}$, assuming detection over $N \times N$ spatial modes (pixels). The parameters $r$ and $\gamma$ represent the relative star-to-exoplanet intensity and the noise floor level, respectively.}
 \label{tab:tab_complexity}
\end{table}

\section{Implementation}

\begin{figure}[t]
    \centering
    \includegraphics[width=\columnwidth,keepaspectratio]{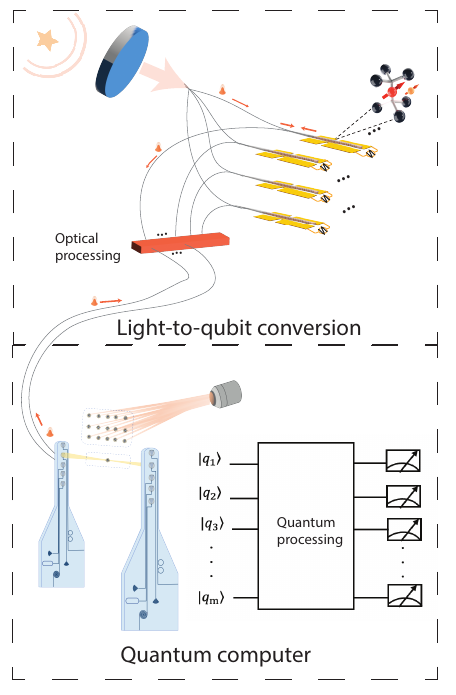}
    \caption{\justifying Sketch of a quantum processing enhanced imaging system implementation. (top) The quantum state of the light collected through the optics is mapped to a SiV-based pixel-qubit register in a heralded way by means of reflection based spin-photon controlled gates followed by an optical quantum Fourier transform and single photon detection (optical processing). (bottom) The quantum state can subsequently be transferred to an atom array quantum processor where our algorithm is applied to separate the light for estimation of individual properties of unresolved sources. The state transfer can be enabled by quantum teleportation through optically mediated entanglement between the atoms and the SiV receivers.  }
   \label{fig:figure1}
\end{figure}

The key functionality of the spin-photon entangling gate needed to coherently map the photonic amplitude information to a qubit register has already been experimentally demonstrated with cavity-coupled Silicon vacancy (SiV) systems~\cite{Bhaskar2020,Knaut2024}. In particular, measurement of the differential phase between two spatially separate stations of weak incident light was recently demonstrated using a network of SiV systems~\cite{telescopeexp}. However extended quantum processing capabilities have not yet been demonstrated with this platform. For practically relevant near term demonstrations of quantum processing sensing, we therefore propose a hybrid implementation utilizing mature quantum computing platforms based such as atomic array qubits~\cite{Bluvenstein2024,rodriguez2024} (see Fig.~\ref{fig:figure1}). 

The optical signal is first collected into an array of single mode optical fibers which routes the light to the SiV based quantum pixels. \jb{Telescope-to-fiber interconnects based on photonic lanterns have already been developed in the context of photon counting receivers with efficiencies of tens of percent~\cite{Jennifer2023}. However, we note that direct free-space coupling to nanophotonic cavity arrays have also been demonstrated with similar efficiencies~\cite{Menon2024}, which could be an alternative implementation.} 

The reflected light can subsequently be combined using optical beam splitters and phase shifters to implement a quantum Fourier transform before detection. This implements the mapping of the photonic amplitude information into the electronic spin qubits of the SiV systems. Photon loss from coupling of light into optical fibers and into the diamond nano-cavities, potential quantum frequency conversion to the resonance frequency of 737nm of the SiV, the optical circuit for the quantum Fourier transformation, and the single photon detection will reduce the achievable gain from the quantum processing detection. However, frequency conversion efficiencies of $\sim 50\%$~\cite{leent2022,stolk2024} have been achieved as well photon-to-qubit gates with efficiencies of $\sim 45\%$~\cite{Bhaskar2020}. Noise in the photon-to-qubit mapping can be filtered by the algorithm as described above at the expense of a lower signal rate since some of the signal is converted to noise by e.g. faulty spin-photon gates.

Access to a second nuclear qubit memory in the SiV systems~\cite{Stas2022,Knaut2024}, can be used to store the received optical information and free up the optically accessible electronic qubit to establish entanglement between the quantum pixel and an atomic processor qubit. The optical information can subsequently be transferred to the processing qubit through quantum teleportation. Notably, current atomic processors have an optical interface used for imaging of the atomic qubits, which can be leveraged for atom-photon entanglement generation. Quantum frequency conversion based on non-linear crystals~\cite{Knaut2024,leent2022}, can be employed to match the frequencies of the diamond color centers (737nm for SiV) to that of the atoms/ions (780 nm for $^{87}$Rb atoms/493 nm for ($^{138}\text{Ba}^+$). In particular, this can be done in a two-step process of upconverting from visible to the telecom band and then down-converting again to the visible to minimize noise from the pump lasers. This allows to benefit from the efficient, broadband ($\sim 1$ GHz) spin-photon interface of Silicon Vacancy centers for encoding of the optical signal into qubits as well as the excellent processing capabilities of ions or atoms.  

The depth and circuit sizes required for our algorithm are very modest compared to other envisioned applications of quantum computing~\cite{Lee2021,beverland2022,dalzell2023}. We require $N^2$ pixel-qubits for capturing the quantum state of an incoming photon across an $N \times N$ pixel array. The information can then be compressed to a qubit register of $\lceil2\log(N)\rceil$ qubits with a total two-qubit gate count of $N^2\log(N)$. The maximum qubit register size required for the quantum processing is set by the requirements to estimate the off diagonal elements in Eq.~(\ref{eq:mes_O}), where a register of $5\lceil2\log(N)\rceil+1$ qubits are needed to store the photonic eigenstates and perform a single qubit controlled SWAP test (see appendix~\ref{app:mes_OBS}). The total number of gates required for the processing is determined by the total number of controlled multi-qubit gates ($e^{-i\epsilon\mathcal{S}}$), which scales as $O(\log(1/\varepsilon_\textrm{st})^2/\varepsilon_\textrm{st})$. Decomposing this into a primitive two-qubit gate count results in an additional factor of $\sim\lceil2\log(N)\rceil$~\cite{Lloyd2014, kimmel2017}.

From the analysis above, we find that a processor with around 36 memory qubits is sufficient to store and process light from a 10×10 pixel array. Both the compression of the 100 pixel-qubits into memory qubits and the PSF sorting algorithm can be implemented using on the order of hundreds of gates for a target SNR ($1/\varepsilon_\textrm{st}$) of 10. Assuming we are imaging in a regime where, with the aid of a coronagraph or starshade, the star's intensity at the detector is approximately an order of magnitude higher than that of the exoplanet, our analysis of the noisy detection indicates errors of around $1\%$ per pixel qubit are sufficient for the photon-to-qubit mapping. A conservative estimate of the required two-qubit gate fidelity can be made by assuming that a single gate error completely corrupts the computation. In this case, the maximum acceptable gate error is approximately the inverse of the total number of gates, divided by the SNR. For $N=10$ and SNR of 10, this corresponds to a per-gate error threshold in the range of $10^{-4}$ to $10^{-3}$. These conservative error estimates and qubit counts are already achievable with current hardware, even without quantum error correction. However, to further reduce noise from imperfect gate operations, the physical states of the SiV qubits could be encoded into modest-sized logical qubits using quantum error correction~\cite{Horsman2012}. 

\jb{Finally, we note that while we intentionally studied a simple depolarizing channel noise model in which the SiV pixel relaxes to a completely mixed state within the single excitation manifold, since this provides an unbiased and hardware independent model, a full device specific error analysis for a specific implementation could further refine the error budgets beyond the conservative bounds estimated above. However, such an analysis is beyond the scope of the present work, since our aim is to evaluate sampling complexity under a broad and conservative noise model.} 

\section{Conclusions and Outlook}

In conclusion, we have introduced a novel technique of quantum processing enhanced optical imaging. The crux of this technique is to coherently transfer the amplitude information of a weak optical field to a qubit processor, which allows to process the signal through qubit gates before detection. Focusing on astronomical imaging, we have shown how a combination of quantum principle component analysis, quantum signal processing and block encoding can be used to sort a weak optical signal consisting of an incoherent mixture of light from two unresolved sources (star and exoplanet) as well as background noise into the (unknown) eigenmodes of the signal. The direct measurement of the eigenmodes allows one to estimate observables of each source individually with a sampling complexity that is independent of the dimension of the system (number of pixels). This lowers the sampling complexity by several orders of magnitude for typical imaging conditions compared to classical tomographic techniques where the sampling complexity always scales with the dimension of the system. 

While we have explicitly considered the setting of two unresolved sources consistent with a single star and exoplanet system, our technique generalizes to more sources. Similarly to the case of noisy detection considered above, the QSP sequence can be used to sort the incoming light into the eigenmodes of multiple sources through a binary search provided that the eigenvalues are distinct. However, diagonalization of the density matrix can not in general be performed analytically for more than two sources and would have to be performed numerically to determine the decomposition of the expected value of the observable from different sources into expected values in the eigenbasis. The eigenvalues can be estimated directly from the sampling while additional information from multiple reference observables and prior information about relative properties of the sources may be necessary to determine off-diagonal terms in the decomposition. The required number of reference observables as well as degree of prior information about relative properties of the sources will depend on the exact signal and observable. We leave it to future work to apply our technique to more complex signals and observables of interest. The achievable reduction in sampling complexity compared to classical tomographic techniques will also depend on the exact form of the signal as also evident from our analysis above of noise-free vs. noisy detection. 

We have outlined how the required functionality of photon-to-qubit mapping can be achieved with existing hardware based on diamond color centers through spin-dependent reflection from a cavity-spin system. Additionally, the information can subsequently be transferred to mature quantum processing platforms based on ions or atoms through optically mediated entanglement between the two types of hardware. Quantum frequency conversion based on non-linear crystals can be used to match the optical resonances of the different hardware. The same frequency conversion can be employed to map signals from a broad range of optical frequencies into the diamond color centers

The bandwidth of cavity-coupled SiVs of $\sim1$ GHz is in the broader end of typical quantum memories. However, it is arguably narrow compared to typical astronomical detection bandwidths which can be a factor of $10^3-10^4$ times broader. Detecting with a broader bandwidth increases the amount of signal collected and hence the SNR. Frequency multiplexing, where the incoming light is frequency de-multiplexed into multiple $\sim 1$ GHz bands can be combined with the quantum detector to increase the detection bandwidth. Each $\sim 1$ GHz band can be matched to an SiV system through frequency conversion allowing to effectively increase the bandwidth of the detection while keeping the high frequency resolution of $\sim$1 GHz. Finally, we note that e.g. studies of rotational broadening of molecular lines from key biosignatures require narrow band detections~\cite{Spring2022} for which the quantum processing-enhanced detection is particularly suited. 

While we have focused on the task of exoplanet imaging, we envision that the concept of quantum processing enhanced imaging can be applied to a number of similar imaging tasks. In particular, the full programability of a quantum processor allows to implement general unitaries to the signal before detection similar to how adaptive optics implements wavefront transformations to denoise the signal before detection. Applying quantum processing enhanced sensing to other areas such as molecular imaging~\cite{Liu2015} and satellite detection and monitoring~\cite{Choi2024} are interesting directions for future work.   

\begin{acknowledgments} 
We thank Breann Sitarski, Saikat Guha, Viatcheslav Dobrovitski, and Soonwon Choi for valuable discussions and insights related to this work. We gratefully acknowledge support from The AWS Quantum Discovery Fund at the Harvard Quantum Initiative, the National Science Foundation (Grant No. PHY-2012023 and  OMA-2326787), NSF Engineering Research Center for Quantum Networks (Grant No. EEC-1941583) and the Center for Ultracold Atoms (Grant No. PHY-2317134), an NSF Frontier Center.
\end{acknowledgments}

\section*{Data availability}
Code for generating the plots is publicly available \cite{QSPastroImage}.

\bibliography{ref}

\appendix

\section{\label{app:psf_model}PSF description}

In full generality, the transformations of propagating light waves are described by Maxwell's equations. However, solving these equations directly is challenging, motivating the use of reduced parametric models. In practice, a telescope—along with the free-space region between the source and the observation point—can often be effectively modeled as a black-box optical system \cite{Goodman}. This system has three key planes: an object plane, a pupil plane, and a detector plane. All three planes are orthogonal to the optical axis of the telescope. We assume that the scientific object is a compact combination of objects, such as a planetary system or a system of several stars, so the optical path through space is almost the same. The distance to the object plane denoted  as $z_o$. The objects are projected onto the object plane and have two real coordinates. We consider objects of relatively compact size, such as stars and planets. Due to their scale, they can be interpreted as point sources emitting spherical electromagnetic waves. By the time these waves reach the observation point, they have an enormous radius and thus appear as nearly plane waves determined by the direction of propagation. The pupil plane contains a device that apodizes the plane waves and applies a phase mask. Afterwards, the waves go through one more free propagation region inside the telescope with the propagation distance $z_i$. On the detector plane located pixelated quantum sensor. We assume that we have a wavelength filter with very narrow broadband, so we are operating the light of the same wavelength $\lambda$. Photons transmitted through the telescope subjected to distortions which shape their phase and amplitude distributions. Functional dependence may be represented as a complex valued function $H(u , v, \xi, \nu)$ describing complex amplitude over the detector plane, coordinates $(u,v)$, of a photon emmited at the point  $(\xi, \nu)$ of the object plane. We refer to $H$ as the complex point spread function (PSF), unlike classical detection, where typically the PSF term is attributed to $|H|^2$, with the quantum processor we are manipulating with the complex quantum state up to the final measurement, hence we don't assume the reduction to the intensity distribution. Analyzing optical path and applying relevant diffraction model the PSF is described as a parametrically dependent integral kernel, those parameters are determined by data-driven approach. One point source at $(\xi_f, \nu_f)$ gives the ``portrait" of the noise on the given optical path $\psi(u,v) = {H} (u , v, \xi_f, \nu_f)$. This portrait captures properties of the noise in the optical system at the time moment, when it was taken. Usual assumption that for two point sources in the same small neighborhood, known as isoplanatic region, the spread functions are given by the same convolution kernel so this functions are just shifts of one function $\psi_k(u,v) =  H(u-\xi_{f_k},v - \nu_{f_k}), \, k=1..2$. On the wide field of view (FOV) PSF may be seen as a family of such functions, for example, it may be given for some covering of the full FOV with isoplanatic regions, where on each isoplanatic region the local PSF has the same parametric form, with parameters depending on region properties, for example, on coordinates of a region's center. Modelling of this PSF variations on a wide FOV is the big topic in the field of astronomical imaging~\cite{Perrin2012,piotrowski2013,Liaudat2023,Redding2024}. If the accurate model of the PSF is known, appropriate processing helps to restore the true image of the scientific object.

Total transformation of the electromagnetic field emitted by the point source at the coordinate $\mathrm{r}_0=(\xi,\nu,z_{o})$ are given by
\begin{equation}\label{eq:analytic_state}
\begin{split}
&\psi(u,v, \mathrm{r}_0) = \frac{1}{\lambda^2 z_0 z_{int}} \exp\left(i \frac{k(\xi^2 + \nu^2)}{2z_{o}} \right)  \\ 
&\exp\left(i \frac{k(u^2 + v^2)}{2z_{i}} \right)  \int_{\mathbb{R}^2} \mathcal{P} (x,y) \exp \left( i \Phi(x,y)\right)\\
&\exp\left( -ik\left(\frac{ u x}{z_{i}} + \frac{v y }{z_{i}} \right) \right) 
\exp\left( -ik\left(\frac{ \xi x}{z_{o}} + \frac{\nu y }{z_{o}} \right) \right) dx dy,
\end{split}
\end{equation}
for multiple sources the field is given as a sum of separate sources. In this expression $\mathcal{P} (x,y): \mathbb{R}^2 \rightarrow \mathbb{R}$ is a real valued function of the pupil plane coordinate, called pupil function, and $\Phi(x,y):\mathbb{R}^2\rightarrow [-\pi, \pi)$ is a phase. Integration performed over the pupil plane, but $\mathcal{P} (x,y)$ always has a compact support determined by the aperture mask of the telescope. We are using Fraunhofer diffraction model in the free propagation regions. This description leads to an expression of the light field on the detector plane emitted by the source located at the point $(\xi, \nu)$
\begin{equation}\label{eq:thePSF}
\begin{split}
&{H} (u , v, \xi, \nu) = \frac{1}{\lambda^2 z_0 z_{int}} \exp\left(i \frac{k(\xi^2 + \nu^2)}{2z_{o}} \right) \\
&\exp\left(i \frac{k(u^2 + v^2)}{2z_{i}} \right) K\left(\frac{u}{z_i} + \frac{\xi}{z_o}; \frac{v}{z_i} + \frac{\nu}{z_o}\right),
\end{split}
\end{equation}
where the function $K$ is a two dimensional Fourier transform of the complex pupil function 
\begin{equation}\label{eq:OTF}
K(p,q) = \mathcal{F}\left[\mathcal{P} (\cdot,\cdot) \exp \left( i \Phi(\cdot,\cdot)\right)\right]\left(\frac{k}{2\pi}p, \frac{k}{2\pi}q\right).
\end{equation}

\section{\label{app:SWAP_test}Swap test}
The SWAP test is a quantum circuit that allows estimating the overlap of two states \cite{barenco1997, Buhrman2001}. Consider two pure states, $\ket{\alpha}$ and $\ket{\beta}$. One additional ancillary qubit is initially in the state $\frac{1}{\sqrt{2}}\left(\ket{0} + \omega\ket{1} \right)$, where $|\omega| = 1$. Applying the controlled-$\textrm{SWAP}$ gate and the Hadamard gate on the ancilla, we prepare the state:
\begin{equation}\label{eq:SWAP_test_state}
\begin{split}
SW_{\omega} \ket{\alpha, \beta}= 
\frac{1}{2}\ket{0}\left( \ket{\alpha, \beta}  +  \omega\ket{\beta, \alpha}\right) \\
+ \frac{1}{2}\ket{1}\left( \ket{\alpha, \beta}  - \omega\ket{\beta, \alpha}\right).
\end{split}
\end{equation}

Let $SW_{\omega,0 (1)}(\ket{\alpha}, \ket{\beta})$ denote the post-measurement state of the register if the ancilla is measured to $0$ ($1$). We denote the measurement record of the ancilla as $SW_{\omega, M}$. For the $\omega = 1$ the probability of obtaining zero is
\begin{equation}\label{eq:SWAP_test_statistics}
\mathbb{P}\left(SW_{1,M}(\ket{\alpha}, \ket{\beta}) = 0\right) = \frac{1}{2} +  \frac{1}{2} \left|\braket{\alpha}{\beta}\right|^2.
\end{equation}

This procedure provides a highly effective method for the important problem of exoplanet detection. Where is an exoplanet when $\rho$ is mixed. The value of $r$ can be measured using the SWAP test for two pairs of $\rho$:
\begin{equation}\label{eq:postSWstate}
\begin{split}
&SW_{1}(\rho, \rho) = r^2 \Bigg[\ket{V_1, V_1}\bra{V_1, V_1}  + \\
&(1-r) r \left(\ket{V_1, V_2} + \ket{V_2, V_1}\right) 
\left(\bra{V_1, V_2} + \bra{V_2, V_1}\right)\\ 
&+ (1-r)^2 \ket{V_2, V_2}\bra{V_2, V_2} \Bigg] \ket{0}\\
&+ (1-r) r \Bigg[\left(\ket{V_2, V_1} - \ket{V_1, V_2}\right) \left(\bra{V_2, V_1} - \bra{V_1, V_2}\right) \Bigg]\ket{1}.
\end{split}
\end{equation}

The statistic of the SWAP test is expressed as follows:
\begin{equation}\label{SWtest_clean_prob} 
\begin{split}
\mathbb{P}\left(SW(\rho, \rho) = 0\right) = 1 - r + r^2.
\end{split}
\end{equation}
Interpreting the experiment as a Bernoulli scheme with two outcomes provides a statistical test that achieves the same statistical error as the B-SPADE quantum detection procedure \cite{lu2018}, which is known to be optimal for this problem, but is based on knowledge of the telescope's PSF function model. Furthermore, the SWAP test is robust to additive noise at the same level as B-SPADE.

\section{\label{app:mes_b}Measuring $b$}
The incoming state $\rho$ has two special representations as mixtures of pure states. 
The first representation is determined by the pair of non-orthogonal signal states $\ket{\psi_1}$ and $\ket{\psi_2}$, while the second representation is given by the pair of eigenvectors. The mixing coefficient $b$ is often assumed to be known. To determine this coefficient, it is necessary to establish a relationship between some values measured separately on $\ket{\psi_1}$ and $\ket{\psi_2}$. Without such a relationship, no measurement can distinguish one pair of states from the continuum of pairs that decompose $\rho$ into convex combinations. The expectation of the observable $\mathrm{O}$ in the state $\Psi$ we denote as $M_{O,\Psi} = \mathrm{Tr}(\mathrm{O} \Psi)$. The relation may have a form of the functional dependence between expectation of one observable
\begin{equation}\label{eq:eq_cond_psi}
\begin{split}
&M_{O_{F},\psi_2} = F\left(M_{O_{F},\psi_1}\right),
\end{split}
\end{equation}
where $F$ is some known function. The observable can be chosen by theoretical, numerical and experimental characterization of the telescope. In addition, we can measure eigenvalues with the SWAP test and choose the relation depending on the outcome. Measuring $\mathrm{O}_{\textrm{F}}$ on $\rho$ we get
\begin{equation}\label{eq:b_eq1}
\begin{split}
&bM_{O_{F},\psi_1} + (1-b)F\left(M_{O_{F},\psi_1}\right) =M_{O_{F},\rho}
\end{split}
\end{equation}

The coefficients $c_{1,k}, c_{2,k}$ of $\ket{\psi_k}, k = 1, 2$, in the basis $\ket{V_k}, k = 1, 2$ have the expressions in terms of the variables $b$ and $h$ (\ref{eq:v_k_eigenv}). These coefficients are real and nonzero. Variable $h$ could be eleminated using (\ref{eq:determ_rel}). Additional equation on $M_{O_{F},\psi_1}$ and $b$ we get measuring on $\ket{V_k}$,
\begin{equation} \label{eq:mix_mes_eq}
\begin{split}
&M_{O_{F},V_k} = |c_{1,k}|^2 M_{O_{F},\psi_1} + \\
&|c_{2,k}|^2F\left(M_{O_{F},\psi_1}\right)+2c_{1,k}c_{2,k} M_c,
\end{split}
\end{equation}
where 
\begin{equation} \label{eq:off_diag_re}
M_c = \Re{\bra{\psi_1}O_{F}\ket{\psi_2}}, 
\end{equation}
eliminating this value from the two last equations we are getting system of two non-linear equations for $b$ and $M_{O_{F},\psi_1}$. The system is non-trivial if at least one condition holds true
\begin{equation} \label{eq:non_triv_cond_mes_b}
M_c \neq 0  \,\,\,  \text{or}\,\, M_{O_{F},\psi_1} \neq M_{O_{F},\psi_2}.
\end{equation}
Then the system can be solved numerically.

\section{\label{app:mes_O_from_ref} Measuring overlaps using reference}

Consider the state
\begin{equation} \label{eq:plusVstate}
\begin{split}
 \ket{W_{\omega}} = \frac{1}{\sqrt{2}} \left(\ket{V_1 V_2} + \omega \ket{V_2 V_1}\right),
\end{split}
\end{equation}
where $\omega$ is unimodular. By measuring the observable $\mathrm{O}_{\text{ref}} \otimes \mathrm{O}$ in this state, we can estimate the off-diagonal term for $\mathrm{O}$. Define 
\begin{equation} \label{eq:kappas}
\begin{split}
\kappa_{\text{ref}}& = \bra{V_1} \mathrm{O}_{\text{ref}} \ket{V_2},\\
\kappa_{\text{int}}& = \bra{V_1} \mathrm{O} \ket{V_2}.
\end{split}
\end{equation}
Measuring $\mathrm{O}_{\text{ref}} \otimes \mathrm{O}$ yields
\begin{equation} \label{eq:O_ref_mes}
\begin{split}
&2\bra{W_{\omega}} \mathrm{O}_{\text{ref}} \otimes \mathrm{O}\ket{W_{\omega}} = \omega\kappa_{\text{ref}}\overline{\kappa_{\text{int}}} + \overline{ \omega}\overline{\kappa_{\text{ref}}}\kappa_{\text{int}} +\\
&\bra{V_1} \mathrm{O}_{\text{ref}} \ket{V_1}\bra{V_2} \mathrm{O} \ket{V_2} + \bra{V_1} \mathrm{O} \ket{V_1}\bra{V_2} \mathrm{O}_{\text{ref}} \ket{V_2} =\\
&\mathfrak{h}_{\omega} + T.
\end{split}
\end{equation}
By measuring this quantity for $\omega = 1$ and $\omega = i$, we obtain
\begin{equation} \label{eq:kappa_int_bar_sys}
\begin{split}
\Re{\kappa_{\text{ref}}\overline{\kappa_{\text{int}}}} = \mathfrak{h}_{1}/2, \quad
\Im{\kappa_{\text{ref}}\overline{\kappa_{\text{int}}}} = -\mathfrak{h}_{i}/2.
\end{split}
\end{equation}
Finally,
\begin{equation} \label{eq:kappa_formula}
\kappa_{\text{int}}
= \frac{\mathfrak{h}_{1} + i\,\mathfrak{h}_{i}}{2\,\overline{\kappa_{\mathrm{ref}}}}.
\end{equation}

Creating a state like (\ref{eq:plusVstate}) requires a copy of $\ket{V_2}$, which can be inefficient for sampling. Therefore, we use indirect approach. First, we directly measure $\bra{V_1}O \ket{V_1}$. To measure the quantity in (\ref{eq:O_ref_mes}), we employ the SWAP test circuit $SW_{\omega}$ (\ref{eq:SWAP_test_state}). For $\omega = i$, we apply $SW_{i}$ to the state $\ket{V_1}$ and $\rho$:
\begin{equation} \label{eq:swi_rhoV1}
\begin{split}
&SW_{i,0}(\ket{V_1},\rho) = r \ket{V_1V_1}\bra{V_1V_1} +(1-r)\ket{W_{i}}\bra{W_{i}},\\
&SW_{i,1}(\ket{V_1},\rho) = r \ket{V_1V_1}\bra{V_1V_1} +(1-r)\ket{W_{-i}}\bra{W_{-i}}.
\end{split}
\end{equation}
Thus, by measuring $\mathrm{O}_{\text{ref}} \otimes \mathrm{O}$ on these states and subtracting the $\ket{V_1V_1}$ contribution, we obtain the system:
\begin{equation} \label{eq:sys_swi_rhoV1}
\begin{cases}
M_0 = \mathfrak{h}_{i} + T, \\
M_1 = -\mathfrak{h}_{i} + T,
\end{cases}
\end{equation}
where $M_0, M_1$ denote measurement outcomes. From this, we immediately obtain $\mathfrak{h}_{i}$, and using the relation $2T = M_1 + M_0$, we estimate $\bra{V_2} \mathrm{O} \ket{V_2}$.

Finally, to handle the case $\omega = 1$, we apply the circuit $SW_{1}$ to the pair of states $\rho$:
\begin{equation} \label{eq:sw1_rhorho}
\begin{split}
SW_{1,j}(\rho,\rho) = r^2 \ket{V_1V_1}\bra{V_1V_1} +\\
(1-r)^2 \ket{V_2V_2}\bra{V_2V_2} + 2r(1-r) \ket{W_{-1^{j}}}\bra{W_{-1^{j}}}.
\end{split}
\end{equation}
Measuring $\mathrm{O}_{\text{ref}} \otimes \mathrm{O}$, with post-processing conditional on ancilla measurement $j$, provides a sample of $\mathfrak{h}_{1}$. Thus, using this protocol, we estimate all terms in (\ref{eq:mes_O}).

\section{\label{app:mes_OBS} Measuring off-diagonal terms using block-encodings}

Consider the observable $\mathrm{O}_{\text{ref}}$ acting on the register with a pure-state Hilbert space $H$ of the same dimension as our main signal state $\rho$. We require access to a block-encoded version of $\mathrm{O}_{\text{ref}} \otimes \mathrm{O}_{\text{ref}}$, i.e., a unitary $U_{\mathrm{O}_{\text{ref}}}$ acting on $H \otimes H \otimes H_{\text{anc}}$, where $H_{\text{anc}}$ corresponds to the ancillary register, such that the following holds:
\begin{equation}
   \left(\bra{0} \otimes \mathrm{I}\right)U_{\mathrm{O}_{\text{ref}}}  \left(\ket{0} \otimes \mathrm{I}\right) = \mathrm{O}_{\text{ref}} \otimes \mathrm{O}_{\text{ref}}.
\end{equation}
The observable $\mathrm{O}_{\text{ref}} \otimes \mathrm{O}_{\text{ref}}$ is assumed to satisfy $||\mathrm{O}_{\text{ref}} \otimes \mathrm{O}_{\text{ref}}|| \leq 1$ in the spectral norm; otherwise, it must be normalized. This enables the following transformation:
\begin{equation}
\begin{split}
    &\ket{V_1 V_1} \ket{0} \mapsto  || \mathrm{O}_{\text{ref}}\ket{V_1}||^{-2} \left(\mathrm{O}_{\text{ref}}\ket{V_1} \otimes \mathrm{O}_{\text{ref}}\ket{V_1} \otimes \ket{0}\right) =\\
    &\ket{V_{\mathrm{M}}}\ket{0},
\end{split}
\end{equation}
where the transformation is applied probabilistically with success probability $|| \mathrm{O}_{\text{ref}}\ket{V_1}||^{4}$.

Using the SWAP test circuit, we can create the state $\ket{W_{1}}$, defined in (\ref{eq:plusVstate}). The overlap with the prepared $\ket{V_{\mathrm{M}}}$ can be estimated using the standard SWAP test:
\begin{equation}
\begin{split}
&|\bra{W_{1}}\ket{V_{\mathrm{M}}}| = \frac{1}{\sqrt{2} \, ||\mathrm{O}_{\text{ref}}\ket{V_1}||^{2}} \bigg| \bra{V_1 V_2} \big( \mathrm{O}_{\text{ref}}\ket{V_1} \otimes \mathrm{O}_{\text{ref}}\ket{V_1} \big) +\\
&\bra{V_2 V_1}\big( \mathrm{O}_{\text{ref}}\ket{V_1} \otimes \mathrm{O}_{\text{ref}}\ket{V_1} \big)  \bigg|=\\
&\frac{|\bra{V_1}\mathrm{O}_{\text{ref}}\ket{V_1}|}{\sqrt{2} \, ||\mathrm{O}_{\text{ref}}\ket{V_1}||^{2}} \bigg|\Re \bra{V_1}\mathrm{O}_{\text{ref}}\ket{V_2}\bigg|.
\end{split}
\end{equation}
The prefactor could be evaluated by measuring the corresponding observables on $\ket{V_1}$. Analogously, starting from $\ket{W_{-1}}$, we can estimate $\bigg|\Im \bra{V_1}\mathrm{O}_{\text{ref}}\ket{V_2}\bigg|$. If we have prior information about the sign of these real and imaginary parts, we can determine the full off-diagonal term $\bra{V_1}\mathrm{O}_{\text{ref}}\ket{V_2}$. In our measurement scheme, we need to perform this procedure once for a well-chosen observable for which the probability of success is high and the sign information is available.

\section{\label{app:Noisy_compls}Sampling complexity in the presence of noise}

In the first scheme we apply QSP circuit based on the noisy state~(\ref{eq:rho_2}) twice. Similarly to the calculation in the last paragraph of the Section~\ref{sec:QSP_for_hev} choosing $\Delta = x(1-\gamma) (1-r)$ we conclude that in the first scheme the complexity of the two steps is 
\begin{equation} \label{eq:2seqQSPcompl}
\begin{split}
O\left(\frac{\left[1+(1-\gamma)^{-2} (1-r)^{-2} \right]\log(\delta)^2}{\epsilon}\right). 
\end{split}
\end{equation}
In the second two level scheme complexity depends on the spectral gap in the new state
\begin{equation} \label{eq:V2_V_r_sp_gap}
\tilde{p}_{2} - \tilde{p}_{\textrm{res}} = \frac{(1-\gamma)(1-r)}{ 1 - r + \gamma r  - \frac{\gamma}{N^2}},
\end{equation}
and the spectral gap separating new state from $\ket{V}_1$, which is $2r-1$.
The final complexity in this scheme is the product of the complexities of separate stages
\begin{equation} \label{eq:V2_V_r_compl}
O\left(\frac{\log(\delta)^4 \left( 1 - r + \gamma r  - \frac{\gamma}{N^2}\right)^2}{\epsilon^2  \left(\left(2r-1 \right)   (1-\gamma)(1-r)\right)^2}\right),
\end{equation} 
if $\gamma \lessapprox (1 - r)$ then the big factor $(1-r)^{-1}$ is suppressed.

\section{\label{app:Davis_Kahan} Numerical analysis of the constant in equation of Davis-Kahan theorem}
Our goal is to construct a perturbation of the matrix \eqref{eq:incoming_state} that turns inequality \eqref{eq:davis_Kahan_est} into an equality with the constant $C\gg 0$. Fix the precision $\varepsilon_{\mathrm{tom}} < r$. Let $\ket{V_{\textrm{pert}}}$ be a vector orthogonal to both $\ket{V_{k}}$, and define $\ket{V_{+}} = \frac{\ket{V_{\textrm{pert}}} + \ket{V_{2}}}{\sqrt{2}}$. Consider the following perturbation of $\rho$:
\begin{equation}\label{eq:pert_rho_DH}
\rho_{\textrm{pert}} = (1 - \varepsilon_{\mathrm{tom}}) \rho + \varepsilon_{\mathrm{tom}} \ket{V_{+}}\bra{V_{+}},
\end{equation}
this is a perturbation of desirable order $||\rho_{\textrm{pert}} - \rho|| \propto \varepsilon_{\mathrm{tom}}$. Let $\ket{\overline{V_{2}}}$ denote the perturbed eigenvector corresponding to $\ket{V_2}$. The ratio of the actual precision $||\ket{V_2} - \ket{\overline{V_{2}}}||$ to the theoretical value $\frac{\varepsilon_\mathrm{tom}}{(1-r)}$ is shown in Figure~\ref{fig:devKahanheatmap}. Valid noise values are less than $1-r$ for all such values we observe that the numerical error coincides with the upper bound of the Davis-Kahan theorem. 

\begin{figure}[ht]\label{fig:devKahanheatmap}
    \centering
    \includegraphics[width=\columnwidth,keepaspectratio]{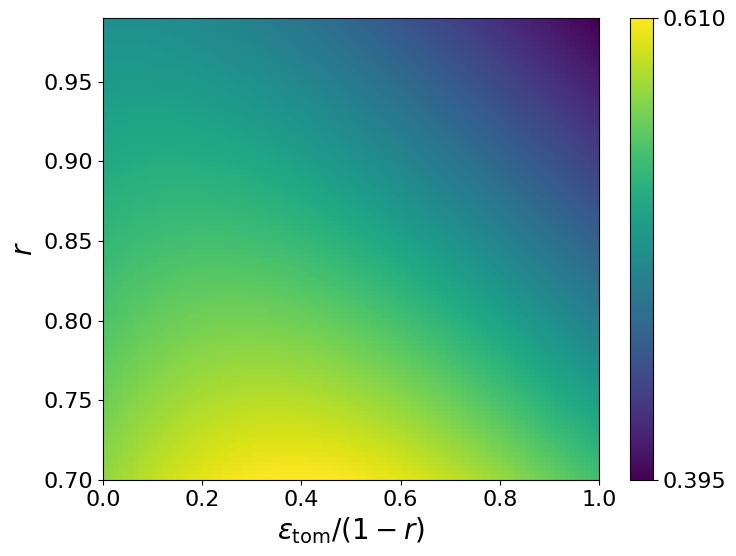}
    \caption{\justifying Ratio of the accuracy of numerical simulation to theoretical prediction, for each $r$ value of $\varepsilon_{\mathrm{tom}}$ varies within the interval $(0, 1 - r)$.}
    \label{fig:devKahanheatmap}
\end{figure}

\end{document}